\documentclass[aps,prd,twocolumn,nofootinbib,preprintnumbers,superscriptaddresss,showkeys,showpacs,amsmath,amssymb,include graphics]{revtex4-1}
\usepackage[a4paper, left=2cm, right=2cm, top=3cm, bottom=4cm]{geometry}
\usepackage{tabularx}
\usepackage{graphicx}
\usepackage{amsmath,amssymb}
\usepackage{amsfonts}
\usepackage{blindtext}
\usepackage{xspace} 
\usepackage[usenames]{color}
\usepackage{dcolumn}
\usepackage{bm}
\usepackage{mathrsfs}
\usepackage[colorlinks=true]{hyperref}
\usepackage[all]{hypcap} 
\usepackage[utf8]{inputenc} 
\usepackage{slashed}
\usepackage{multirow}
\usepackage{rotating}
\usepackage{nicematrix}
\usepackage{makecell}
\usepackage[bbgreekl]{mathbbol}
\usepackage{relsize}
\DeclareSymbolFontAlphabet{\mathbb}{AMSb}
\DeclareSymbolFontAlphabet{\mathbbl}{bbold}
\usepackage{tikz-feynman}
\usepackage{tikz}
\usetikzlibrary{calc}
\usepackage{float}
\usepackage{ragged2e}

\usepackage[normalem]{ulem}

\usetikzlibrary{external}

\begin{document}

\title{Color-superconducting quarkyonic matter}

\author{Christoph Gärtlein} \email{christoph.gartlein@tecnico.ulisboa.pt} 
\affiliation{Centro de Astrof\'{\i}sica e Gravita\c c\~ao  - CENTRA, Departamento de F\'{\i}sica, Instituto Superior T\'ecnico - IST, Universidade de Lisboa - UL, Av. Rovisco Pais 1, 1049-001 Lisboa, Portugal}
\affiliation{CFisUC, Department of Physics, University of Coimbra, 3004-516 Coimbra, Portugal}
\affiliation{Institute of Theoretical Physics, University of Wroclaw, 50-204 Wroclaw, Poland}

\author{Oleksii Ivanytskyi} \email{oleksii.ivanytskyi@uwr.edu.pl}  
\affiliation{Incubator of Scientific Excellence---Centre for Simulations of Superdense Fluids, University of Wrocław, 50-204, Wroclaw, Poland}

\author{Violetta Sagun} 
\email{v.sagun@soton.ac.uk}
\affiliation{Mathematical Sciences and STAG Research Centre, University of Southampton, Southampton SO17 1BJ, United Kingdom}

\author{Ilídio Lopes}
\email{ilidio.lopes@tecnico.ulisboa.pt} 
\affiliation{Centro de Astrof\'{\i}sica e Gravita\c c\~ao  - CENTRA, Departamento de F\'{\i}sica, Instituto Superior T\'ecnico - IST, Universidade de Lisboa - UL, Av. Rovisco Pais 1, 1049-001 Lisboa, Portugal}

\date{\today}

\begin{abstract}
We explore the role of color superconductivity in quarkyonic matter under the conditions of color and electric neutrality at $\beta$- and strong equilibrium, as relevant for neutron stars. By explicitly incorporating the color-superconducting pairing gap into the phenomenological model of a smooth transition from hadron to quark matter, we extend the known quarkyonic framework to include this essential aspect relevant at high densities. The momentum dependence of the pairing gap, motivated by the running of the QCD coupling and introduced similarly to chiral quark models with nonlocal interaction, is a novel element of the model that is crucial for enabling the simultaneous onset of all color–flavor quark states in the presence of color superconductivity. While asymptotically conformal behavior of the present model is ensured by construction, we demonstrate that reaching the conformal limit in agreement with the predictions of perturbative QCD is provided by the proper momentum dependence of the thickness of the hadron shell in momentum space. We employ the flexible meta-modeling approach to nuclear matter, analyzing the structure of the hadron shell in momentum space and focusing on the effects of color superconductivity in quarkyonic matter. Similar to the effects induced by the onset of the quarkyonic phase, color superconductivity leads to stiffening of the equation of state of the NS matter. This causes a significant impact on observable properties of neutron stars, which are analyzed and compared to recent astrophysical and theoretical constraints. We argue that the developed model of color-superconducting quarkyonic matter provides a new, consistent tool for studying the scenario of smooth quark-hadron transition in NSs.

\end{abstract}
\maketitle

\section{Introduction}
\label{intro}


The significant advances in observational astronomy over recent years have made it possible to obtain accurate measurements of various neutron star (NS) characteristics~\cite{Antoniadis:2013pzd,LIGOScientific:2017vwq,LIGOScientific:2018hze,LIGOScientific:2018cki,Miller:2019cac,Riley:2019yda,NANOGrav:2019jur,Miller:2021qha,Riley:2021pdl,Fonseca:2021wxt}. This provides a unique opportunity to probe the properties of extreme matter existing in the interiors of NSs, while neither the terrestrial experiments on collisions of relativistic heavy ions~\cite{Iancu:2012xa,Shuryak:2014zxa,Busza:2018rrf,Dietrich:2020efo,Sorensen:2023zkk} nor numerical simulation in lattice Quantum Chromodynamics (QCD)
~\cite{Bazavov:2017dus,Bazavov:2017dsy,HotQCD:2018pds,Ratti:2018ksb,Borsanyi:2018grb,Borsanyi:2020fev,Guenther:2020jwe,Borsanyi:2020fev,Borsanyi:2021sxv} can access the baryon densities typical for these astrophysical objects. Thus, with the usual masses in the range $1.2$ - $2~\rm M_\odot$ and radii about $11$ - $13$ km, NSs can support the baryon densities several times the nuclear saturation density ~\cite{Baym:2017whm,Vidana:2018lqp,Capano:2019eae,MUSES:2023hyz,Koehn:2024set}.

While deconfinement of quarks, arising from the dissociation of hadrons into their constituents, is an inevitable consequence of the asymptotic freedom of QCD at high densities~\cite{Gross:1973id,Politzer:1973fx,Glendenning:1997wn}, the open question remains {\it whether NSs support such high densities} or, equivalently, {\it whether deconfined quark matter exists in the NS interiors}?

Recent NICER measurements evidence that the radius of heavy PSR J0740+6620 with the mass above $2\rm M_\odot$~\cite{Salmi:2024aum,Dittmann:2024mbo} exceeds the radii of the objects PSR J0437-4715~\cite{Choudhury:2024xbk}, PSR J0030+0451 ~\cite{Vinciguerra:2023qxq,Miller:2019cac} and PSR J0614-3329 \cite{Mauviard:2025dmd} with the intermediate masses about $1.4\rm M_\odot$.
Relatively small radius of the canonical mass NSs is also supported by the gravitational-wave signal GW170817~\cite{LIGOScientific:2017vwq}.
These observational facts can be reconciled if the NS equation of state (EoS) is relatively soft at $2$ - $3$ nuclear saturation densities and significantly stiffens at higher densities due to a strong vector repulsion~\cite{Dexheimer:2020rlp}.
This generates a characteristic back-bending in the mass-radius relation of NSs with masses $1.4$ - $2\rm M_\odot$.
Such back-bending is typical in the presence of quark cores in NSs~\cite{Alvarez-Castillo:2018pve,Contrera:2022tqh,Ivanytskyi:2022oxv,Ivanytskyi:2022bjc,Carlomagno:2023nrc,Blaschke:2023,Gartlein:2023vif,Li:2024lmd,Ivanytskyi:2024zip,Gartlein:2024cbj} but is challenging for most of the purely hadronic EoSs.

According to the reported small radius of the ultralight HESS J1731-347 compact object \cite{Doroshenko2022}, the mentioned back-bending extends even below one solar mass. While the compactness of the HESS J1731-347 is yet debated, the NICER data on PSR J1231-1411~\cite{Salmi:2024bss} suggest rather large radii of NSs with masses about $1\rm M_\odot$ and, consequently, stiff NS EoS at 2 nuclear saturation densities.
Consistency of this second possibility with the mentioned above soft NS EoS at $1$ - $1.4\rm M_\odot$ can be provided by a strong first-order phase transition occurring at those NS masses or even below them.
A strong first-order phase transition at small NS mass can also explain the reported small radius of the HESS J1731-347 object making it consistent with the existence of quark matter in the interiors of all NSs~\cite{Sagun:2023rzp}.
The nature of the Black Widow Pulsar J0952-0607 with the mass $2.35\pm0.17~{\rm M}4_\odot$~\cite{Romani:2022jhd} also can be explained as an NS with color-superconducting (CS) quark core. 
Very recent results of physics-informed Bayesian analysis of observational data also provide strong statistical evidence in favor of the existence of quark cores in NSs compared to purely hadronic NS scenarios~\cite{Ayriyan:2025rub}. In addition, as recently demonstrated~\cite{Sagun:2023rzp}, hybrid stars with paired quark matter in their cores show good agreement with observational data on the thermal evolution of NSs, including HESS J1731-347~\cite{Doroshenko2022}.

If NSs indeed contain quark matter, they provide a unique opportunity to probe the properties of the transition between quark and hadronic matter, which is of fundamental importance for understanding the QCD phase diagram~\cite{Wilczek:1999ym,Rischke:2003mt,Stephanov:2006zvm,Fukushima:2010bq,Sorensen:2023zkk,Murgana:2025wsh,Abuali:2025tbd,Li:2025ugv,Schmidt:2025ppy,INDRA:2025htq}.
In the case of a sharp transition between hadronic and quark degrees of freedom, the phase transition is of the first order, characterized by a discontinuous jump in density. Usually, such a phase transition is modeled by the Maxwell construction~\cite{Baym:2017whm}.
This is equivalent to the Gibbs criterion of phase equilibrium under the condition of baryon charge conservation, while the conservation of electric charge is loosened, which leads to a discontinuity of the corresponding chemical potential.
In the case of two or more conserved charges, as when the conservation of electric charge is accounted for along with the conservation of baryon charge, the Gibbs criterion leads to a continuous behavior of the density across the quark-hadron phase transition. This corresponds to the Glendenning construction of the phase transition~\cite{Glendenning:1992vb}, which is the limiting case of the phase coexistence with the zero surface tension between the phases. A more realistic description requires accounting for spatial inhomogeneities, the so-called pasta phases (see, e.g., Ref.~\cite{Maslov:2018ghi}), which are mimicked by various interpolation schemes~\cite{Ayriyan:2017nby,Ayriyan:2021prr,Ivanytskyi:2022wln,Gao:2025vdc}. 

Answering the question about the details of the quark-hadron transition and its relevance for NSs requires a unified quark-hadron approach. The picture of quarkyonic (QY) matter provides a physically transparent framework for this approach~\cite{McLerran:2007qj,Hidaka:2008yy}. Its main phenomenological consequence is a smooth character of the deconfinement phase transition that leads to a pronounced peak of the speed of sound~\cite{McLerran:2018hbz}. This feature has been utilized in modeling the NS interiors~\cite{McLerran:2018hbz,Zhao:2020dvu,Zhang:2020jmb,Park:2021hqb,Koch:2022act,Cao:2022inx,Pang:2023dqj,Poberezhnyuk:2023rct,Duarte:2023cki,Gao:2024jlp,Folias:2024upz,Dey:2024lco,Gao:2024jlp,Fujimoto:2024doc,Fujimoto:2025sxx,Dey:2025jbm,Bluhm:2024uhj} reaching the high masses consistent with observational constraints~\cite{Antoniadis:2013pzd,Fonseca:2021wxt}. Recently, the QY picture has also been applied for modeling cold QCD matter at vanishing baryon densities and large isospin asymmetries~\cite{Ivanytskyi:2025cnn}.

QY matter might give rise to distinct dynamical signatures, for instance, in binary NS mergers. While a sharp deconfinement phase transition generates resonant tidal excitations of interfacial i-modes during the binary NS coalescence~\cite{Counsell:2025hcv}, QY matter might produce a different effect. The smooth onset of quarks does not create a distinct interface, thereby making its gravitational wave imprint 
very different from the one of a first-order quark-hadron transition with a systematic decrease of the dominant postmerger frequency ~\cite{Bauswein:2018bma}. This effect is caused by vanishing incompressibility of the mixed phase at a first-order phase transition and is not observed in the case of a smooth mixed phase modeled within the Glendenning construction~\cite{Prakash:2021wpz} or with an interpolation scheme~\cite{Kedia:2022nns}. Relative stiffening of a smooth mixed phase region compared to the sharp one leads to prolonging the postmerger stage but does not significantly modify the dominant postmerger frequency compared to the purely hadronic scenario. Based on this we can expect that significant stiffness of the QY matter right after the onset of quark matter (see e.g.~\cite{McLerran:2018hbz}) can prolong the postmerger phase even more and increase its dominant frequency.

While resembling the scenario of quark-hadron continuity~\cite{Schafer:1998ef,Baym:2017whm,Kojo:2021ugu,Fujimoto:2023mzy} and being supported by holographic models~\cite{Kovensky:2020xif,Yang:2020hun,Chen:2019rez}, the QY picture is based on the assumption that quarks remain bound to hadrons up to the quark chemical potentials $\mu_q$ being parametrically
large compared to the QCD energy scale $\Lambda$.
The 't Hooft limit of a large number of colors $N_c$ and the vanishing QCD coupling as $g_{\rm QCD}\propto1/N_c$ allow for a simple illustration of this picture.
The confining interactions are suppressed when the Debye mass exceeds the QCD energy scale. 
At zero temperature typical for the NS interiors~\cite{Fukushima:2010bq,Kurkela:2009gj,Baym:2017whm}, the Debye mass scales as $g_{\rm QCD}^2N_c\mu_q^2$. 
With this, we conclude that the confining interactions remain important up to $\mu_q \simeq\sqrt{N_c}\Lambda$\footnote{In other words, the deconfinement chemical potential being parametrically large compared to the QCD scale assumes the scaling $\mu_B/\Lambda\simeq\sqrt{N_c}$, which is large in the 't Hooft limit.}. 
This binds quarks from the vicinity of their Fermi sphere of size $k_h$ to hadrons. 
The momentum states resided deeply inside of the Fermi sea and having momenta smaller than $k_q$ do not experience the confinement since the Pauli blocking among them disables the confining scatterings~\cite{McLerran:2007qj,Hidaka:2008yy}.
Thus, the QY picture suggests that quarks with $k_q\le|{\bf k}|\le k_h$ are confined to hadrons, while the states with $|{\bf k}|<k_q$ remain effectively unbound. At small densities, $k_q$ vani\-shes so there are no unbound quarks, while at high densities $k_q$ approaches $k_h$ and hadrons disappear due to their dissociation to the constituent quarks.

The described QY picture is similar to the phenomenon of Cooper pairing, which also mostly affects the high-momentum states in the vicinity of the Fermi sphere~\cite{Cooper:1956zz,Bardeen:1957mv}. Accounting for this and for the possible interpretation of baryons as color singlet bound states of quarks and diquarks ~\cite{Praschifka:1986nf,Cahill:1988zi,Cahill:1988bh,Burden:1988dt,Reinhardt:1989rw,Zuckert:1996nu,Wang:2010iu,Blanquier:2011zz}, we conclude that diquarks should also play a role in QY matter where baryons are constituted by the quark states from the vicinity of the Fermi sphere.
Bose-Einstein condensation of diquarks corresponds to the formation of CS~\cite{Son:1998uk,Alford:1999pa,Abuki:2003ut,Yuan:2024ajk}, which plays an important role in the phenomenology of NSs~\cite{Buballa:2003qv,Baym:2017whm}.
It is worth mentioning that since superconductivity is inevitable for any sufficiently dense fermionic system with an arbitrarily weak attractive interaction~\cite{Cooper:1956zz}, the CS should necessarily manifest itself in QY matter.
However, to the best of our knowledge, CS has never been accounted for within the framework of QY matter. The present paper fills this gap, introducing a model of CS quarkyonic (CSQY) matter in the case of two-flavor CS (2CS)~\cite{Buballa:2003qv,Baym:2017whm}.
We also model NSs with CSQY cores and show that this scenario provides a good agreement with the observational data.

The paper is organized as follows. In Sec.~\ref{sec2}, we introduce the single-particle distribution functions of quarks in CSQY matter as well as the corresponding expressions for the number and energy densities. The general formulation of the model of CSQY matter, the conditions of strong equilibrium, color neutrality, $\beta$-equilibrium, and electric neutrality typical for NSs are presented in Sec.~\ref{sec3}. Section~\ref{sec4} is devoted to modeling NSs with CSQY cores. The conclusions are given in Sec.~\ref{sec5}. Throughout the paper, we work in the natural units with $\hbar=c=G=1$.

\begin{figure}[t]
\begin{minipage}{\linewidth}
\includegraphics[width=\columnwidth]{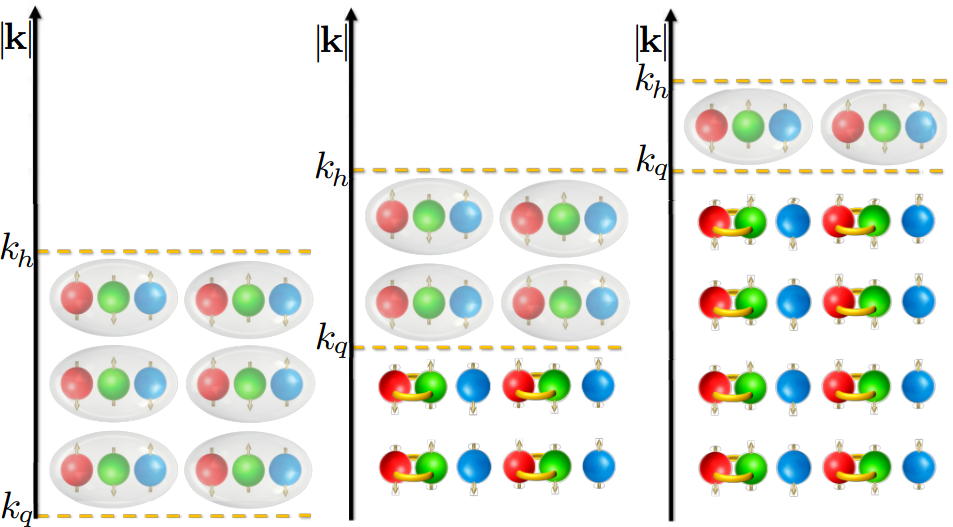}
\end{minipage}
\caption{Schematic illustration of CSQY matter in momentum space. The orange dashed lines indicate the maximum momentum of quarks confined in hadrons $k_h$ and the Fermi momentum of unbound quarks $k_q$. The red, green, and blue spheres represent quarks of the corresponding color, with the spins indicated by the arrows. The flavor states are not indicated. Gray glass-like spheres depict three quarks confined to color-singlet nucleons of two possible spins. Dimming the red, green, and blue quarks by these gray spheres depicts confinement of colored degrees of freedom.
The golden rings represent scalar diquarks formed by the Cooper pairing of deconfined red and green quarks, while deconfined blue quarks below the quark Fermi sphere remain unpaired. At small densities below the onset of QY matter (left panel), all quarks remain bound to hadrons, and quark Fermi momentum vanishes.
At intermediate densities above the onset of QY matter (middle panel), unbound quarks form the Fermi sphere with finite $k_q$, while the quarks bound to nucleons exist in the momentum shell between $k_q$ and $k_h$. At high densities (right panel) the width of this shell vanishes.}
\label{fig:art}
\end{figure}
%

\section{Quarks in Color-Superconducting Quarkyonic matter}
\label{sec2}

In the presence of the 2SC phase, only two color states of quarks are paired, while the third one is not~\cite{Buballa:2003qv}. Conventionally, we define the red and green color states to be paired, while the blue color state is unpaired. 
The phenomenological picture of CSQY matter consistent with this convention and with the notations introduced in Sec. \ref{intro} is shown in Fig. \ref{fig:art}.
All the color states of quarks are labeled by the subscript index $c=(r,g,b)$. Thus, below we quantify the effects of the 2SC phase by the color-dependent pairing gap $\Delta_{c{\bf k}}=(\Delta_{\bf k},\Delta_{\bf k},0)$. The momentum dependence of the gap $\Delta_{\bf k}$ is motivated by the running of the QCD coupling~\cite{Deur:2016tte} and is introduced similarly to the nonlocal Nambu–Jona–Lasinio (NJL) model~\cite{Blaschke:2022egm,Contrera:2022tqh,Carlomagno:2023nrc,Ivanytskyi:2024zip}.

The color dependence of the 2SC matter pairing gap leads to an inequivalence among quarks of different colors, resulting in distinct Fermi momenta $k_{fc}$ that generally do not coincide\footnote{\color{blue} The mismatch of quark momenta in the presence of CS is discussed in the review \cite{Schmitt:2025cqi}.}. 
Hereafter, the index $f=(u,d)$ labels quark flavor states with masses $m_f$.
In comparison to the chiral quark models, in which the medium-dependent quark masses are defined within a self-consistent mean-field approximation (see e.g. Refs.~\cite{Buballa:2003qv,Ivanytskyi:2022oxv,Ivanytskyi:2024zip}), for the sake of simplicity, we treat the quark masses as constant parameters of the model.
We also allow them to deviate from the current quark masses from the Review of Particle Physics~\cite{ParticleDataGroup:2022pth} and assume that they partially account for the complexity of quark in-medium interactions. As is shown below, the Fermi energy of a given flavor-color state of quarks is $\epsilon_{fc}-\Delta_{fc}$, where $\Delta_{fc}$ denotes the pairing gap defined at the quark Fermi momentum and $\epsilon_{fc}=\sqrt{k_{fc}^2+m_f^2}$.

The single-particle distribution function of unpaired blue quarks is given by the Fermi-Dirac distribution controlled by the corresponding Fermi momentum. The single-particle distribution of paired quarks can be derived within the Nambu-Gorkov formalism~\cite{Nambu:1960tm,Gorkov:1958}. However, this formalism treats quarks of all momenta as unbound quasiparticles, while the striking element of the QY picture is that the momentum states above the Fermi sphere are bound to baryons~\cite{McLerran:2018hbz}. This circumstance can be accounted for by cutting off from the Nambu-Gorkov single-particle distribution the momentum states above $k_{fc}$.

To define the single-particle distribution function of quarks at a given color-flavor state in CSQY matter, we start with the corresponding Nambu-Gorkov number density~\cite{Buballa:2003qv,Baym:2017whm,Ivanytskyi:2022oxv,Ivanytskyi:2024zip}
\begin{eqnarray}
    \label{I}
    \tilde{n}_{fc}=2\sum_\pm\int\frac{d{\bf k}}{(2\pi)^3}
    \left[\frac{1}{2}-\theta(-\epsilon^\pm_{fc{\bf k}})\right]
    \frac{\epsilon_{fc}\pm\epsilon_{f{\bf k}}}{\epsilon^\pm_{fc{\bf k}}}.
\end{eqnarray}
Hereafter, the tilded quantities are obtained without cutting off the quark states above the Fermi surface. The factor $2$ in Eq. (\ref{I}) stands for the spin degeneracy of quarks; the summation is performed over the particle and antiparticle states. Eq. (\ref{I}) includes the single-particle energy shifted by $\epsilon_{fc}$ for particles or its negative for antiparticles, i.e.
\begin{eqnarray}
    \label{II}
    \epsilon^\pm_{fc{\bf k}}={\rm sgn}(\epsilon_{f{\bf k}}\pm\epsilon_{fc})
    \sqrt{(\epsilon_{f{\bf k}}\pm\epsilon_{fc})^2+\Delta^2_{c{\bf k}}},
\end{eqnarray}
where $\epsilon_{f{\bf k}}=\sqrt{{\bf k}^2+m_f^2}$. Considering this expression right below the Fermi sphere of quarks, we conclude that the quark Fermi energy, indeed, is $\epsilon_{fc}-\Delta_{fc}$.

The first term in the squared brackets in Eq. (\ref{I}) accounts for the zero-point oscillations of quarks. In the absence of paring ($\Delta_{c{\bf k}}=0$) it vanishes. The second term accounts for the particle excitations below the Fermi sphere with $\epsilon_{f{\bf k}}<\epsilon_{fc}$. The zero-point term can be rewritten using the identity $1/\epsilon_{fc{\bf k}}^\pm=[\theta(\epsilon_{fc{\bf k}}^\pm)-\theta(-\epsilon_{fc{\bf k}}^\pm)]/|\epsilon_{fc{\bf k}}^\pm|$.
This allows us to customize the Nambu-Gorkov number density as
\begin{eqnarray}
    \label{III}
    \tilde{n}_{fc}=2\int\frac{d{\bf k}}{(2\pi)^3}
    \tilde{f}_{fc{\bf k}},
\end{eqnarray}
where the single-particle distribution of paired quarks, which ignores their confinement to baryons above the Fermi sphere, is
\begin{eqnarray}
    \label{IV}
    \tilde{f}_{fc{\bf k}}
    =\frac{1}{2}
    \sum_\pm\frac{\epsilon_{fc}\pm\epsilon_{f{\bf k}}}{|\epsilon_{fc{\bf k}}^\pm|}.
\end{eqnarray}
Remarkably, this distribution function explicitly includes a contribution of antiparticles, which does not vanish even in the zero-temperature case. It appears due to the nontrivial structure of the zero-point term caused by CS, which leads to the redistribution of the quark momentum space.

\begin{figure}[t]
\includegraphics[width=1\columnwidth]{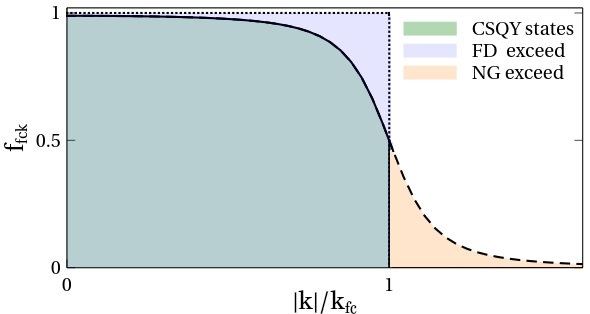}
\caption{The single-particle distribution function of paired quarks $f_{fc{\bf k}}$ as a function of their momentum given in units of the Fermi momentum, i.e. $|{\bf k}|/k_{fc}$. The dotted, dashed, and solid curves correspond to the Fermi-Dirac distribution of unpaired quarks, the Nambu-Gorkov distribution of paired quarks with the high momentum states above the Fermi sphere, and the distribution of paired quarks in CSQY matter with a cut-off of the high momentum tail, respectively. The CSQY and Nambu-Gorkov distributions are calculated for constant $\Delta_{c{\bf k}}=0.15\epsilon_{fc}$. The green, light blue, and light orange shaded areas represent the occupation of the quark states in CSQY matter; the Fermi-Dirac exceeds these states below the Fermi sphere, and the Nambu-Gorkov exceeds above it.}
\label{fig1}
\end{figure}

At a vanishing pairing gap, the Nambu-Gorkov distribution function (\ref{IV}) reduces to the zero-temperature Fermi-Dirac distribution $\theta(\epsilon_{fc}-\epsilon_{f{\bf k}})$. At a finite pairing gap below the Fermi surface $\tilde{f}_{fc{\bf k}}<\theta(\epsilon_{fc}-\epsilon_{f{\bf k}})=1$ and $\tilde{f}_{fc{\bf k}}>\theta(\epsilon_{fc}-\epsilon_{f{\bf k}})=0$ above it. This manifests in the redistribution of the quark momentum states mentioned above (see Fig.~\ref{fig1}).
According to the QY picture, the momentum states above the Fermi sphere (orange area in Fig. \ref{fig1}) are no longer part of the single particle spectrum of free quarks but redistribute to the spectrum of baryons.
Therefore, these states should be cut off from $\tilde{f}_{fc{\bf k}}$ and consistently accounted for through the distribution function of baryons.
This produces the single-particle distribution of quarks in CSQY matter. 
Thus, we define
\begin{eqnarray}
    \label{V}
    f_{fc{\bf k}}
    =\frac{\theta(\epsilon_{fc}-\epsilon_{f{\bf k}})}{2}
    \sum_\pm\frac{\epsilon_{fc}\pm\epsilon_{f{\bf k}}}{|\epsilon_{fc{\bf k}}^\pm|}.
\end{eqnarray}
In the absence of CS, the distribution function (\ref{V}) also converges to the zero-temperature Fermi distribution. Fig.~\ref{fig1} shows it along with the Nambu-Gorkov and Fermi-Dirac distributions. As is seen, quark pairing in CSQY matter reduces their distribution function compared to the case when the pairing is absent, while all the momentum states up to the Fermi one remain filled. As a result, the quark number density in CSQY matter 
\begin{eqnarray}
    \label{VI}
    n_{fc}=2\int\frac{d{\bf k}}{(2\pi)^3}f_{fc{\bf k}},
\end{eqnarray}
gets reduced compared to the corresponding density of normal QY matter. This effect is reflected in the CS induced stiffening of the QY matter\footnote{The review~\cite{Baym:2017whm} provides an excellent discussion of stiffness of quark-hadron EoSs.}, which is discussed in Sec.~\ref{sec4}.

In the absence of CS, the antiparticle contribution in the single-particle distribution function of quarks is absent. 
In this case, the energy density of quarks attains the form of a momentum integral of the pro\-duct of the Fermi-Dirac distribution function and single-particle energy at vanishing pairing gap. In the case of CSQY matter, the contribution of antiparticles does not va\-nish, and each term of this distribution should be multiplied by the corresponding single-particle energy not shifted by the quark Fermi energy, i.e., by $\epsilon^\pm_{fc{\bf k}}\mp\epsilon_{fc}$. The resulting energy density reads
\begin{eqnarray}
    \label{VII}
    \varepsilon_{fc}=\sum_\pm\int\frac{d{\bf k}}{(2\pi)^3}
    \theta(\epsilon_{fc}-\epsilon_{f{\bf k}})
    \frac{\epsilon_{fc}\pm\epsilon_{f{\bf k}}}{|\epsilon_{fc{\bf k}}^\pm|}(\epsilon^\pm_{fc{\bf k}}\mp\epsilon_{fc}).\nonumber\\
\end{eqnarray}
Note, the spin degeneracy factor of quarks in this expression is canceled by the factor $1/2$ from the expression for their single-particle distribution in CSQY matter. At vanishing pairing gap, Eq. (\ref{VII}) converges to the energy density corresponding to the Fermi-Dirac distribution of quarks.

The described scheme of accounting for CS in QY matter by modifying the Nambu-Gorkov distribution of paired quarks suggests a reasonable compromise between the physical transparency of QY matter and the technical complexity of its complete microscopic theory in the presence of CS. 
Such a theory can be constructed similarly to the QY picture of isospin QCD~\cite{Ivanytskyi:2025cnn}. 
This requires a self-consistent determination of the momentum shell of bound quarks and a dynamical evaluation of the spectral properties of nucleons, which are considered as bound states of quarks in the mentioned momentum shell.

\section{Color-Superconducting Quarkyonic matter}
\label{sec3}

In this section, we formulate a model of CSQY matter. In addition to two flavors of quarks, the model also includes nucleons, labeled by the subscript index $a=(n,p)$, and electrons. Below, we label all the particle states with the subscript index ``$i$'' running over $a$ for nucleons, $fc$ for quarks, and $e$ for electrons. In Subsec.~\ref{subsec3a}, we present the general formulation of the model and derive its most important thermodynamic quantities. The conditions of equilibrium with respect to strong decays of nucleons to their constituent quarks (strong equilibrium) and color neutrality are considered in Subsec.~\ref{subsec3b}. Subsec.~\ref{subsec3c} is devoted to imposing the conditions of electric neutrality and $\beta$-equilibrium, which are important for modeling NSs.

\subsection{General formulation of the model}
\label{subsec3a}

Similarly to normal QY matter, in CSQY matter nucleons occupy the momentum states in the shell of width $\delta_a$ above the quark Fermi momentum. 
In principle, this width should be self-consistently determined by relating it to the number of quark states above the Fermi sphere, as has been done in the regime of isospin QCD~\cite{Ivanytskyi:2025cnn}. 
However, this requires accounting for the quark substructure of nucleons manifested by the dynamically generated medium-dependence of their mass.
For the sake of simplicity, in this work we omit this step and, following Ref.~\cite{McLerran:2018hbz}, parametrize the width of the nucleon shell as a piecewise power function of their highest momentum $k_a$, i.e.,
\begin{eqnarray}
    \label{VIII}
    \delta_a=k_a\theta(\Lambda_a-k_a)
    +\frac{\Lambda_a^{\kappa+1}}{k_a^\kappa}\theta(k_a-\Lambda_a),
\end{eqnarray}
where $\Lambda_a$ is a constant parameter of the order of the QCD energy scale $\Lambda$ and $\kappa$ is a positive integer. In Refs.~\cite{McLerran:2018hbz,Zhao:2020dvu}, a value of $\kappa = 2$ was adopted, albeit with parameter choices different from those used here. In contrast to previous parametrizations of the hadronic shell, we present a new perspective on this topic and explore a different range of $\kappa$ values. At $k_a<\Lambda_a$, the lowest nucleon momentum is zero and, consequently, the corresponding momentum space is nothing but the Fermi sphere with $k_a$ being the Fermi momentum of nucleons. At $k_a>\Lambda_a$, the low-momentum states are not filled by nucleons, which allows populating these states with quarks. Thus, the CSQY matter onsets when the Fermi momentum of nucleons equals $\Lambda_a$. Below, the quantities defined at the onset of CSQY matter are labeled with the superscript index ``$\rm onset$''. For example, the baryon density of the CSQY matter onset is denoted as $n_B^{\rm onset}$. This allows us to fix
\begin{eqnarray}
    \label{IX}
    \Lambda_a=k_a^{\rm onset},
\end{eqnarray}
where $k_a^{\rm onset}$ can be found from the model of nuclear matter described below if $n_B^{\rm onset}$ is specified. The values of $\Lambda_a$ defined in this way depend on the isospin asymmetry of CSQY matter and do not coincide in the general case. This provides the simultaneous appearance of low-momentum cavities in the distribution functions of neutrons and protons.

These distribution functions, along with the number and energy densities of nucleons, read
\begin{eqnarray}
    \label{X}
    f_{a{\bf k}}&=&\theta(|{\bf k}|-k_a+\delta_a)\theta(k_a-|{\bf k}|),\\
    \label{XI}
    n_a&=&2\int\frac{d{\bf k}}{(2\pi)^3}f_{a{\bf k}},\\
    \label{XII}
    \varepsilon_a&=&2\int\frac{d{\bf k}}{(2\pi)^3}f_{a{\bf k}}
    E_{\bf k},
\end{eqnarray}
where the factor $2$ corresponds to the two spin states of nucleons. Their single-particle energy $E_{\bf k}=\sqrt{{\bf k}^2+M^2}$ includes the flavor-blind nucleon mass $M$. Below, the single-particle energy evaluated at $k_a$ is denoted as $E_a$.

Eq. (\ref{XII}) accounts for the kinetic and mass terms of the energy of nucleons. The potential energy is ge\-ne\-ra\-ted by the medium-dependent interactions. Thus, the nuclear component of the CSQY matter energy density can be written down as
\begin{eqnarray}
\label{XIII}
\varepsilon_N=
\sum_a\varepsilon_a+U_{\rm is}+I_N^2U_{\rm iv}.
\end{eqnarray}
The second term in this expression stands for the isoscalar interaction potential, which depends only on the baryon density carried by nucleons $n_N=n_n+n_p$. The third term is the isovector interaction potential in the parabolic approximation. Hereafter the nuclear asymmetry parameter $I_N=(n_nI_n+n_pI_p)/n_N$ is introduced through $I_n=1$ and $I_p=-1$. Note, $U_{\rm iv}$ only depends on $n_N$. The specific form of the potentials $U_{\rm is}$ and $U_{\rm iv}$ used in this work is discussed below.

The total baryon density of CSQY matter is
\begin{eqnarray}
    \label{XIV}
    n_B=n_N+\frac{1}{N_c}\sum_{fc}n_{fc},
\end{eqnarray}
where the first term in this expression represents nucleons, while the second one accounts for the contribution of quarks with baryon charge $1/N_c$. It is clear that $n_B^{\rm onset}=n_N^{\rm onset}$, since quark number densities va\-nish at the onset of CSQY matter.

At high densities, the asymptotic freedom of QCD unbinds quarks from nucleons~\cite{Gross:1973id,Politzer:1973fx}, which leads to the vanishing of nucleon density. Considering that at high maximum momenta of nucleons $k_a$, the nucleon shell in the momentum space gets thinner with a width vanishing $\propto1/k_a^\kappa$, we conclude that at high densities $n_a\simeq\Lambda_a^{\kappa+1}k_a^{2-\kappa}/\pi^2$ in the leading order. At $\kappa=2$ used in Refs.~\cite{McLerran:2018hbz,Zhao:2020dvu} the nucleon densities saturate to the constant values $\Lambda_a^3/\pi^2$, while the nucleon fractions vanish $\propto1/n_B$. Requiring $n_a$ to vanish at high densities in agreement with the asymptotic freedom of QCD, in this work, we consider $\kappa \ge 3$.

The total energy density of CSQY matter also includes contributions from nucleons and quarks. For simplicity, we ignore all the interactions among quarks except the pairing one leading to the 2SC phase discussed above.
Thus, we define
\begin{eqnarray}
\label{XV}
\varepsilon=\varepsilon_N+\sum_{fc}\varepsilon_{fc},
\end{eqnarray}
where the first term corresponds to the nuclear component and the second term represents the contribution of quarks (similar to Eq.(\ref{XIV})).

The chemical potentials of a particle of sort $i$ can be found as the partial derivative of the total energy density with respect to the corresponding number density $n_i$ evaluated at constant values of the number densities of other components $n_{i'\neq i}=const$, i.e.,
\begin{eqnarray}
    \label{XVIa}
    \mu_i=\frac{\partial\varepsilon}{\partial n_i}.
\end{eqnarray}
At the same time, we notice that number densities and Fermi momenta of nucleons enter only the energy density carried by the nuclear component of CSQY matter. With this, we obtain the following expression
\begin{eqnarray}
    \label{XVI}
\mu_a=\frac{\partial\varepsilon_N}{\partial n_a}
\hspace*{-.1cm}&=&\hspace*{-.1cm}
\frac{\partial\varepsilon_a}{\partial k_a}\Bigl/\frac{\partial n_a}{\partial k_a}
\nonumber\\
\hspace*{-.1cm}&+&\hspace*{-.1cm}
\frac{\partial U_{\rm is}}{\partial n_N}
+2I_N(I_a-I_N)\frac{U_{\rm iv}}{n_N}+I_N^2\frac{\partial U_{\rm iv}}{\partial n_N}.\nonumber\\
\hspace*{.6cm}
\end{eqnarray}
The first term in this expression accounts for the mass and kinetic energy of nucleons within the shell. The second one is generated by the isospin-sensitive nuclear interaction, while the last two terms in Eq. (\ref{XVI}) are produced by the isovector part of these interactions. As expected, $\mu_n-\mu_p=E_n-E_p+4I_NU_{\rm iv}/n_N$ is positive in the neutron rich nuclear matter ($I_N>0$) and negative in the proton rich case ($I_N<0$).

The quark Fermi momenta enter only the partial contributions to the total energy density, which yields
\begin{eqnarray}
\label{XVII}
\mu_{fc}=\frac{\partial\varepsilon_{fc}}{\partial k_{fc}}
\Bigl/\frac{\partial n_{fc}}{\partial k_{fc}}.
\end{eqnarray}
Despite the expression for $\mu_{fc}$ can be explicitly found using Eqs. (\ref{VI} - \ref{VII}), it is rather extended. Therefore, we omit it for the sake of shortening the notations. At a vanishing paring gap $\mu_{fc}=\epsilon_{fc}$.

Having the chemical potentials of nucleons and quarks defined, the pressure of CSQY matter can be found using the well-known thermodynamic identity as
\begin{eqnarray}
    \label{XVIII}
    p=\sum_a\mu_a n_a+\sum_{fc}\mu_{fc}n_{fc}-\varepsilon.
\end{eqnarray}
Below we also consider the speed of sound, which is defined as $c_S^2=dp/d\varepsilon$, and the dimensionless interaction measure $\delta=1/3-p/\varepsilon$.
It is worth mentioning that determining the chemical potentials and pressure from the corresponding thermodynamic identities ensures the thermodynamic consistency of our approach.

In this work, we use the minimal parameterization of the isoscalar nuclear potentials $U_{\rm is}$, which is consistent with the properties of the nuclear matter ground state, i.e., at the nuclear saturation density $n_0$ and vanishing isospin asymmetry. Particularly, the vanishing pressure at $n_0$, baryon chemical potential $\mu_0=M-E_B/A$ expressed through the binding energy per nucleon $E_B/A$ and nuclear incompressibility factor $K_0=9n_0^2d^2(\varepsilon_0/n_0)/dn_0^2$. Hereafter, the quantities defined at the nuclear matter ground state are labeled with the subscript index ``0''. We also account for the nuclear skewness coefficient $Q_0=27n_0^3d^3(\varepsilon_0/n_0)/dn_0^3$, which allows us to modify the stiffness of the nuclear EoS at high densities. To reproduce the present data from nuclear physics and astrophysical constraints, we follow the strategy of a meta-modeling approach~\cite{Margueron:2017eqc,Margueron:2017lup, Zhang:2018vrx}, a semi-agnostic, flexible model based on a Taylor expansion of the energy per particle in terms of the particle number density of nucleons and the isospin asymmetry.

Thus, to reproduce the normal nuclear matter properties, we expand the isoscalar nuclear potential around the nuclear saturation density. This expansion is truncated at the cubic term, which allows us to unambiguously fix the corresponding expansion coefficients according to the values of $\mu_0$, $K_0$, and $Q_0$.
Thus,
\begin{eqnarray}
    \label{XIX}
    U_{\rm is}&=&\mu_0n_0-\sum_a\varepsilon_{a0}
              +(\mu_0-E_0)(n_N-n_0)\nonumber\\
              &+&\left(\frac{K_0}{3}-\frac{k_0^2}{E_0}\right)
              \frac{(n_N-n_0)^2}{6n_0}\nonumber\\
              &+&\left(\frac{Q_0}{9}+K_0+\frac{k_0^2}{3E_0}+\frac{k_0^4}{3E_0^3}\right)
              \frac{(n_N-n_0)^3}{18n_0^2},\nonumber\\
\end{eqnarray}
where $k_0=(3\pi^2n_0/2)^{1/3}$ and $E_0=\sqrt{M^2+k_0^2}$ are the Fermi momentum and Fermi energy of nucleons in the nuclear matter ground state with the corresponding density $n_0$. In Sec.~\ref{sec4} and Appendix~\ref{app:A}, we discuss the impact of the cubic term in the expression for $U_{\rm is}$. While the structure of this potential coincides with the one in the nuclear meta-model, Eq. (\ref{XIX}) also explicitly accounts for the contribution originating from the kinetic energy of nucleons $\sum_a\varepsilon_{a0}$.

The nuclear isovector potential $U_{\rm iv}$ is treated simi\-larly; we expand it around the saturation density. However, we truncate the expansion to the second order since it is enough to reproduce the most important parameters of the isospin-sensitive nuclear interaction, which are the nuclear symmetry energy $J_0$ and its slope $L_0$ at the saturation density. For this, we first define the nuclear symmetry energy as half of the second derivative of the nuclear energy per particle with respect to the asymmetry parameter at its zero value, i.e., $J=\sum_ak_a^2/12E_a+U_{\rm iv}/n_N$. This expression gives direct access to the symmetry energy slope $L=3n_N\partial J/\partial n_N$ and includes the derivative of the ratio $U_{\rm iv}/n_N$ with respect to $n_N$. Thus, fixing the values of the symmetry energy and its slope at the saturation density allows us to express the ratio $U_{\rm iv}/n_N$ as a linear function of the baryon density carried by nucleons.
With this, we arrive at
\begin{eqnarray}
\frac{U_{\rm iv}}{n_N}&=&J_0-\frac{k_0^2}{6E_0}\nonumber\\
\label{XX}
&+&\left[L_0-\frac{k_0^2}{6E_0}\left(2-\frac{k_0^2}{E_0^2}\right)\right]\frac{n_N-n_0}{3n_0}.
\end{eqnarray}
This expression corresponds to the second-order expansion of the isovector potential mentioned above.

Below, we utilize the nucleon mass taken as the mean value of the masses of neutron and proton from the Review of Particle Physics~\cite{ParticleDataGroup:2022pth}. The properties of the nuclear matter ground state are fixed according to the analysis of the multi-messenger data, together with the nuclear matter experiments within the meta-modeling framework~\cite{Mondal:2022cva}. 
The chosen symmetry energy slope agrees with the value extracted from the observational data on NSs using a nonparametric Gaussian process EoS~\cite{Tang:2025xib}.
These properties are summarized in Table~\ref{table1}. 
\begin{table}[t]
\centering
\begin{tabular}{|c|c|c|c|c|c|c|}
\hline    
     $M$     & $E_B/A$&     $n_0$      & $K_0$ & $Q_0$ & $J_0$ & $L_0$ \\
$\rm [MeV]$  &  [MeV] & $\rm [fm^{-3}]$& [MeV] & [MeV] & [MeV] & [MeV] \\ \hline
     939     &  16.1  &     0.154      &  230  & -600  &  32   &  45    \\ \hline
\end{tabular}
\caption{Nucleon mass and parameters of the nuclear matter ground state utilized in this work and fixed according to Refs.~\cite{ParticleDataGroup:2022pth,Mondal:2022cva, Scurto:2024ekq}.} 
\label{table1}
\end{table}
To provide agreement with the constraints of subsaturation density of the chiral effective field theory~\cite{Hebeler:2013nza,Drischler:2015eba,Tews:2018kmu}, we matched the nuclear EoS above with the Haensel-Zdunik EoS (HZ) for the outer crust and the Negele-Vautherin (NV) EoS for the inner crust~\cite{1989A&A...222..353H,Negele:1971vb}. The matching is provided utilizing the Maxwell crossing of two EoSs.

\begin{figure}[t]
\includegraphics[width=1\columnwidth]{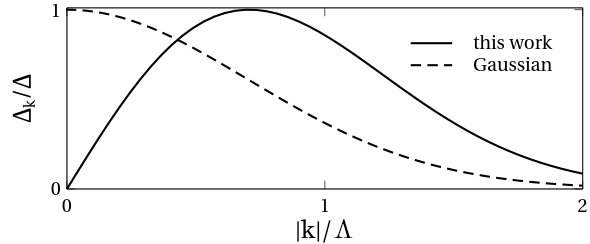}
\caption{Scaled momentum-dependent pairing gap $\Delta_{\bf k}/\Delta$ used in this work (solid curve) compared to the Gaussian one (dashed curve) as functions of the scaled momentum $|{\bf k}|/\Lambda$.}
\label{fig2}
\end{figure}

Before going further, we want to analyze the che\-mi\-cal potentials of quarks at the onset of CSQY matter. At the onset, the quark Fermi momenta va\-nish, leading to the corresponding number and energy densities becoming zero. As a result, the ratio of the derivatives in Eq. (\ref{XVII}) can be replaced by the ratio $\varepsilon_{fc}/n_{fc}$ according to L'H\^opital's rule. To evaluate the number and energy density of quarks at small values of their Fermi momenta, we notice that in this case $\epsilon_{f\bf k}$ and $\epsilon_{fc}$ under the momentum integrals in Eqs. (\ref{VI}) - (\ref{VII}) can be replaced by $m_f$, and the pairing gap can be evaluated at vanishing momentum. With these simplifications we obtain $n_{fc}= m_fk_{fc}^3/2\pi^2\sqrt{(2m_f)^2+(\Delta_{c{\bf k}=0})^2}$ and $\epsilon_{fc}=(\sqrt{(2m_f)^2+(\Delta_{c{\bf k}=0})^2}-m_f)n_{fc}$. This yields
\begin{eqnarray}
    \label{XXI}
    \mu_{fc}^{\rm onset}=\sqrt{(2m_f)^2+(\Delta_{c{\bf k}=0})^2}-m_f.
\end{eqnarray}
Since the simultaneous onset of quarks of three different colors assumes the same value of their chemical potentials, we require the pairing gap to vanish at zero momentum. For this, we define the momentum-dependent pairing gap as
\begin{eqnarray}
    \label{XXII}
    \Delta_{\bf k}=\frac{\sqrt{2}|{\bf k}|}{\Lambda}\exp\left[\frac{1}{2}-\frac{{\bf k}^2}{\Lambda^2}\right]\Delta.
\end{eqnarray}
The low-momentum behavior of $\Delta_{\bf k}$ ensures that the condition $\Delta_{c{\bf k}=0}=0$ is satisfied, leading to $\mu_{fc}=m_f$ at vanishing Fermi momenta of quarks, irrespective of their color. At high momenta $\Delta_{\bf k}$ behaves as the momentum-dependent effective gap in nonlocal versions of chiral quark models with the Gaussian formfactor~\cite{Blaschke:2022egm,Contrera:2022tqh,Carlomagno:2023nrc,Ivanytskyi:2024zip}. The pairing gap (\ref{XXII}) has extremum at $|{\bf k}|=\Lambda/\sqrt{2}$. The gap is defined such that its extremal value is $\Delta$, setting the amplitude of the momentum-dependent pairing gap. Hereafter, we refer to $\Delta$ and $\Delta_{\bf k}/\Delta$ as the pairing gap and the scaled momentum-dependent pairing gap, respectively. Fig.~\ref{fig2} shows behavior of $\Delta_{\bf k}/\Delta$. Compared to the Gaussian case, the parametrization (\ref{XXII}) leads to a weaker pairing of the low-momentum quark states with $|{\bf k}|\lesssim\Lambda/3$ but enhances pairing of the high-momentum states above this momentum. At the same time, in the nonperturbative range of densities typical for the NS interiors, the high-momentum states are suppressed by the single-particle distribution function of paired quarks. Therefore, the mentioned effects compensate each other and, at a given value of $\Delta$, the parametrization (\ref{XXII}) used in this work corresponds to a quark pairing compatible with the Gaussian case.
%
\subsection{Strong equilibrium and color neutrality of color-superconducting quarkyonic matter}
\label{subsec3b}

The QY picture of dense matter assumes a coexistence of nucleons and quarks. This implies establishing a dynamical equilibrium with respect to strong decays of nucleons to their constituents and the opposite processes of recombination. The corresponding conditions are given by the requirement of equality of chemical potentials of nucleons and the sum of the chemical potentials of their decay products. Such conditions were first introduced to the framework of normal QY matter by~\citet{Zhao:2020dvu}. The condition of strong equilibrium for CSQY matter can be formulated as follows. 

Here, we note that nucleons are color singlets and, therefore, are composed of quarks in all three color states. Thus, the flavor-color structure of quarks constituting neutrons is $(u_r,d_g,d_b)$ or $(u_b,d_r,d_g)$ or $(u_g,d_b,d_r)$. Since red and green quark color states are paired with the same pairing gap, they should have the same Fermi momenta and the same chemical potentials. Therefore, the first and the third flavor-color combinations of quarks constituting neutrons correspond to equivalent relations between the chemical potentials. Similarly, among three flavor-color combinations of the quark states corresponding to protons, i.e., $(u_r,u_g,d_b)$, $(u_b,u_r,d_g)$, and $(u_g,u_b,d_r)$, the second and the third ones are equivalent. Therefore, the conditions of strong equilibrium can be formulated as the requirements of equality of $\mu_n$ to the sum of chemical potentials of quarks from the flavor-color combinations $(u_r,d_g,d_b)$ and $(u_b,d_r,d_g)$ as well as equality of $\mu_p$ to the sum of chemical potentials of quarks from the flavor-color combinations $(u_r,u_g,d_b)$ and $(u_b,u_r,d_g)$. These conditions should be supplemented with two requirements $\mu_{fr}=\mu_{fg}$. The mentioned conditions are an inconsistent system of linear algebraic equations with respect to six chemical potentials of quarks.
In other words, this system can not be unambiguously solved if no additional condition is imposed.

Such a condition is naturally provided by the requirement of color neutrality of CSQY matter. This is equivalent to requiring vanishing expectation va\-lues of eight densities $n_a=\langle q^+\lambda_a q\rangle$, with $q$ being a spin-color-flavor spinor of quarks and $\lambda_a$ standing for the generators of the ${\rm SU}(N_c)$ color group given by the Gell-Mann matrices. The index $a$ ranges from $1$ to $N_c^2-1$. Note, the quark number densities are $n_{fc}=\langle q^+\mathcal{P}_{fc}q\rangle$, where $\mathcal{P}_{fc}$ is the projector on a given flavor-color state. In the case of the 2SC phase, all the color densities except $n_8$ vanish automatically. Then, noticing that the eighth Gell-Mann matrix is diagonal and denoting the corresponding matrix elements as $\lambda_{8c}$, the contribution of a given flavor-color state to $n_8$ is obtained as $n_{fc}\lambda_{8c}$.  Thus, the conditions of strong equilibrium and color neutrality can be formulated as
\begin{eqnarray}
    \label{XXIII}
    \mu_n&=&\mu_{ur}+\mu_{dg}+\mu_{db},\\
    \label{XXIV}
    \mu_n&=&\mu_{ub}+\mu_{dr}+\mu_{dg},\\
    \label{XXV}
    \mu_p&=&\mu_{ur}+\mu_{ug}+\mu_{db},\\
    \label{XXVI}
    \mu_{ug}&=&\mu_{ur},\\
    \label{XXVII}
    \mu_{dg}&=&\mu_{dr},\\
    \label{XXVIII}
    \sum_{fc}n_{fc}\lambda_{8c}&=&0.
\end{eqnarray}
Note, the condition of equilibrium with respect to decays of protons to quarks of the flavor-color configu\-ration $(u_b,u_r,d_g)$ is not present here explicitly but is respected implicitly. It can be obtained by subtracting Eq. (\ref{XXIII}) from the sum of Eqs. (\ref{XXIV}) and (\ref{XXV}) with the consequent application of Eqs. (\ref{XXVI}) and (\ref{XXVII}). 

The six conditions imposed by strong equilibrium and color neutrality eliminate six quark Fermi momenta $k_{fc}$ from the set of independent thermodynamic parameters of CSQY matter. The two remaining parameters, i.e., the maximum momenta of nucleons $k_a$, determine the total baryon density and isospin asymmetry of CSQY matter.

As it was shown in Subsec.~\ref{subsec3a}, the chemical potentials of quarks coincide with their mass at the onset of CSQY matter. This allows us to exclude the quark masses from the list of independent parameters of the model. To achieve this, we utilize Eqs. (\ref{XXIV} - \ref{XXV}) at the onset of CSQY matter, which yields
\begin{eqnarray}
    \label{XXIX}
    m_u&=&\frac{2\mu_p^{\rm onset}-\mu_n^{\rm onset}}{3},\\
    \label{XXX}
    m_d&=&\frac{2\mu_n^{\rm onset}-\mu_p^{\rm onset}}{3}.
\end{eqnarray}
These quark masses are isospin-dependent and do not necessarily coincide.

\subsection{$\beta$-equilibrium and electric neutrality of color-superconducting quarkyonic matter}
\label{subsec3c}

The NS matter is electrically neutral and exists in equilibrium with respect to direct and inverse $\beta$-decay processes. To incorporate these features into the developed model of CSQY matter, we follow the minimal scheme described in Ref.~\cite{Zhao:2020dvu} and introduce electrons with mass $m_e=0.511$ MeV~\cite{ParticleDataGroup:2022pth}. Their Fermi momentum $k_e$ is found by requiring a zero value of the total density of electric charge, i.e.,
\begin{eqnarray}
    \label{XXXI}
    n_p+\sum_{fc}n_{fc}Q_f-n_e=0,
\end{eqnarray}
where $Q_f$ is the electric charge of a given quark flavor and $n_e=k_e^3/3\pi^2$ is the number density of electrons. The corresponding energy density and pressure are obtained according to the expressions for massive noninteracting fermions with spin $1/2$. They are added to the total energy density and pressure given by Eqs. (\ref{XV}) and (\ref{XVIII}). The chemical potential of electrons is $\mu_e=\sqrt{k_e^2+m_e^2}$.

The electron Fermi momentum $k_e$ determined by Eq.~(\ref{XXXI}) is a function of two thermodynamic parameters of CSQY matter, which remain independent after imposing the conditions of strong equilibrium and color neutrality, i.e., the Fermi momenta of nucleons. One of them can be found from the condition of $\beta$-equilibrium $\mu_n=\mu_p+\mu_e$. Note, Eqs. (\ref{XXIII}) - (\ref{XXVII}) allow presenting this condition in other equivalent forms. For definiteness, we use it to fix $k_p$, leaving $k_n$ as the only independent thermodynamic parameter of CSQY matter. By varying this parameter, we control the baryon density of the system. 

\section{Neutron Stars with Color-Superconducting Quarkyonic cores}
\label{sec4}

\begin{table}[t]
\centering
\begin{tabular}{|c|c|c|c|c|}
\hline    
$n_B^{\rm onset}$ & $\Lambda_n$ & $\Lambda_p$ & $m_u$ & $m_d$ \\
                  &   [MeV]     &    [MeV]    & [MeV] & [MeV] \\\hline
    $2n_0$        &    403.3    &    165.2    & 237.1 & 403.2 \\\hline 
    $3n_0$        &    459.0    &    202.9    & 255.5  & 459.6 \\\hline
\end{tabular}
\caption{The model parameters defined by the onset density of CSQY matter.} 
\label{table2}
\end{table}

Modeling NSs with CSQY cores requires fixing the parameters of the EoS developed in Sec.~\ref{sec3}. Three of them, i.e., $\kappa$, $\Delta$, and $\Lambda$, are independent, while $\Lambda_n$, $\Lambda_p$, $m_u,$ and $m_d$ are fixed by specifying the onset density of CSQY matter. Below we consider the scenarios with relatively low ($n_B^{\rm onset}=2n_0$) and moderate ($n_B^{\rm onset}=3n_0$) onset density of CSQY matter. This corresponds to the model parameters specified in Table~\ref{table2}. To reduce the parameter space, we also fix $\Lambda=\Lambda_n$. Thus, only $\kappa$ and $\Delta$ remain free, and we systematically vary their values to obtain a set of EoSs, which are used to model NSs with CSQY cores. The presented results are obtained within the model with a rather stiff hadronic EoS, which correctly reproduces the skewness of the isoscalar energy per nucleon. To demonstrate the phenomenological role of this parameter of nuclear matter and to address the uncertainties of the baryonic matter EoS, we perform a similar analysis considering a softer hadronic EoS (for details see Appendix~\ref{app:A}).

It is appropriate to add a brief remark concerning the choice of the onset density of CSQY matter. In general, the QY framework allows any value of $n_B^{\rm onset}$.
However, stiffening of the NS EoS caused by the onset of QY matter (see e.g. Ref.~\cite{McLerran:2018hbz} for discussion) can lead to superluminal $c_S^2$ if $n_B^{\rm onset}$ is high and speed of sound of hadron matter is high itself.
At the same time, having this quantity above the onset density of hyperons, which is about $3n_0$ (see e.g. Refs.~\cite{Yamamoto:2015lwa,Yamamoto:2017wre,Li:2019sxd}), requires accounting for these particles in hadronic EoS.
To provide reliability of the simple phenomenological parameterization of the used hadronic EoS and to omit the necessity to account for hyperons we restrict the onset density of CSQY matter to the values discussed above.
It is necessary to stress that they agree with the results of physics-informed Bayesian analysis of the NS data, suggesting that quarks most probably onset at $2n_0$ - $3n_0$ \cite{Ayriyan:2025rub}.
Recently, a very similar range of $n_B^{\rm onset}$ has been reported within a Bayesian analysis of the observational data on NSs with a nonparametric Gaussian process EoS~\cite{Tang:2025xib}.
While that study also suggests higher onset densities of quarks about $5n_0$, the analysis does not account for hyperons.
Therefore, we follow a more reliable range of $n_B^{\rm onset}\simeq 1n_0-2n_0$, which recently has also been used for constructing an interpolating quark-haron EoS~\cite{Gao:2025vdc}, as a guidance.

\begin{figure*}[ht]
\centering

\begin{minipage}{0.48\textwidth}
    \centering
    \begin{tikzpicture}[remember picture]
        \node[anchor=south west, inner sep=0] (main1) 
        {\includegraphics[width=\linewidth]{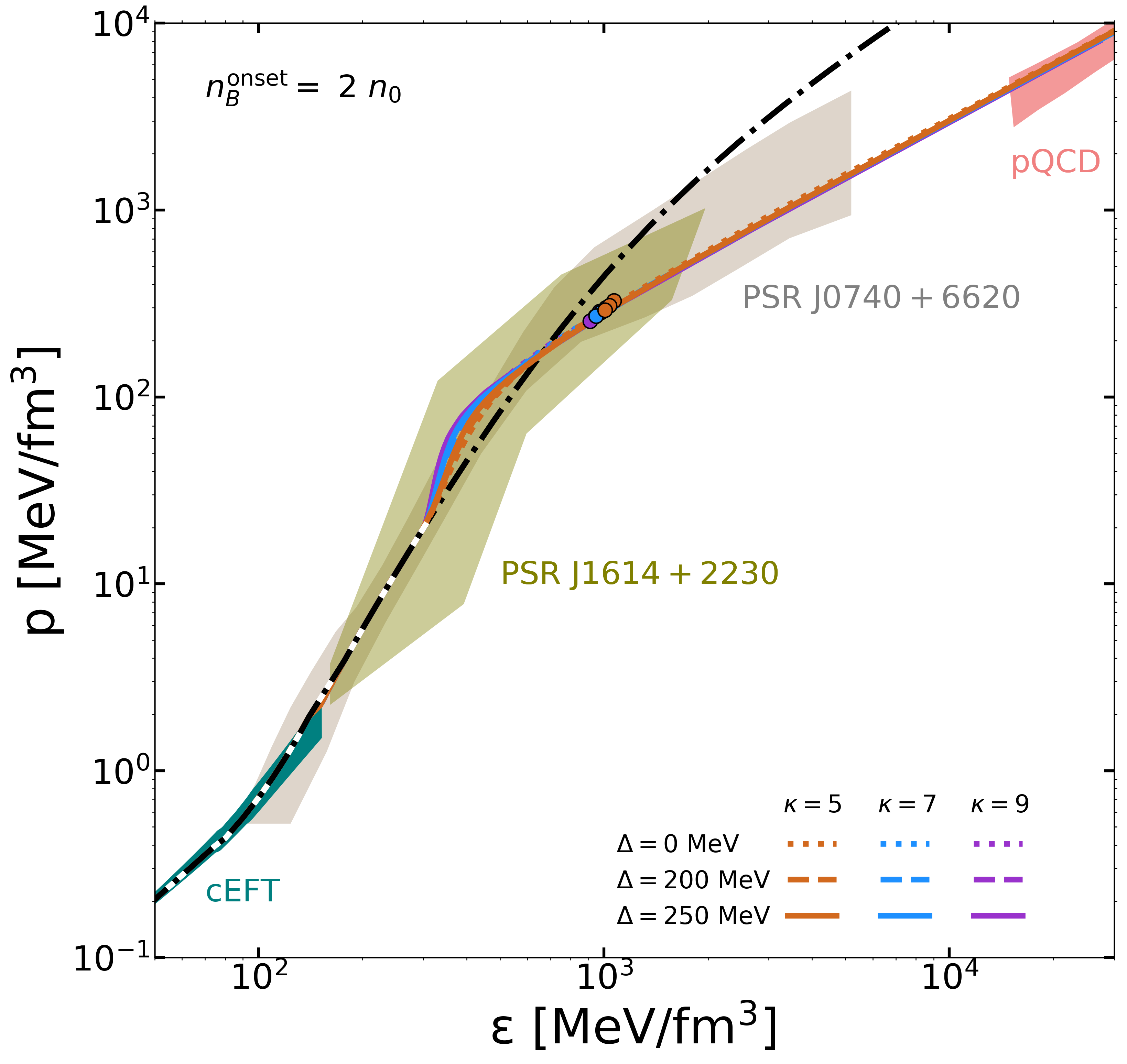}};
        \begin{scope}[x={(main1.south east)}, y={(main1.north west)}]
            \coordinate (r1SW) at (0.35,0.49);
            \coordinate (r1SE) at (0.51,0.49);
            \coordinate (r1NE) at (0.51,0.685);
            \coordinate (r1NW) at (0.35,0.685);
            \draw[gray, very thick] (r1SW) rectangle (r1NE);
        \end{scope}
    \end{tikzpicture}
\end{minipage}
\hfill
\begin{minipage}{0.48\textwidth}
    \centering
    \begin{tikzpicture}[remember picture]
        \node[anchor=south west, inner sep=0] (zoom1) 
        {\includegraphics[width=\linewidth]{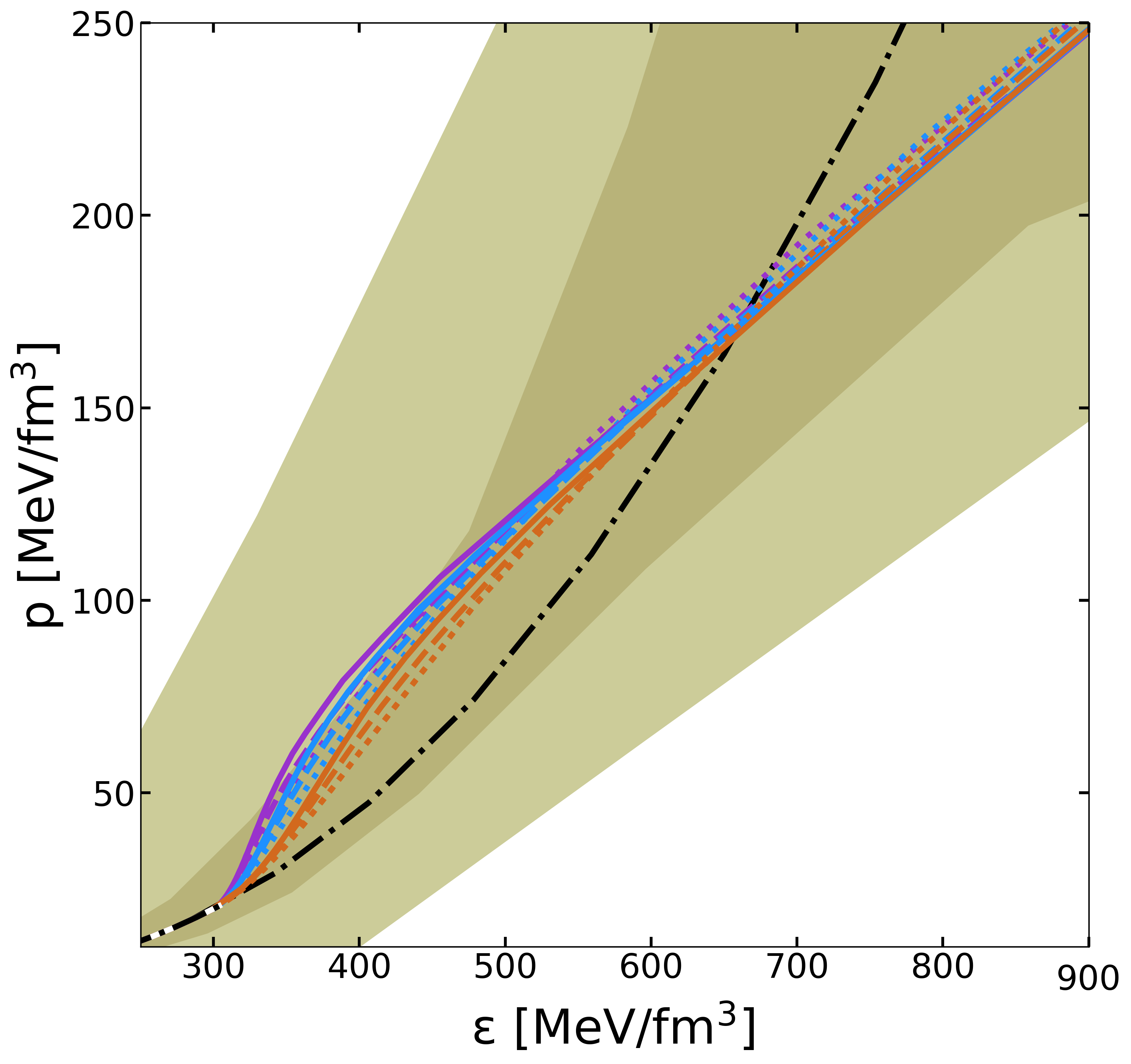}};
        \coordinate (z1NW) at ($(zoom1.north west)+(3em,-0.5em)$);
        \coordinate (z1NE) at ($(zoom1.north east)+(-1em,-0.6em)$);
        \coordinate (z1SW) at ($(zoom1.south west)+(3.2em,2.7em)$);
        \coordinate (z1SE) at ($(zoom1.south east)+(-1em,2.7em)$);
    \end{tikzpicture}
\end{minipage}

\vspace{1em}

\begin{minipage}{0.48\textwidth}
    \centering
    \begin{tikzpicture}[remember picture]
        \node[anchor=south west, inner sep=0] (main2) 
        {\includegraphics[width=\linewidth]{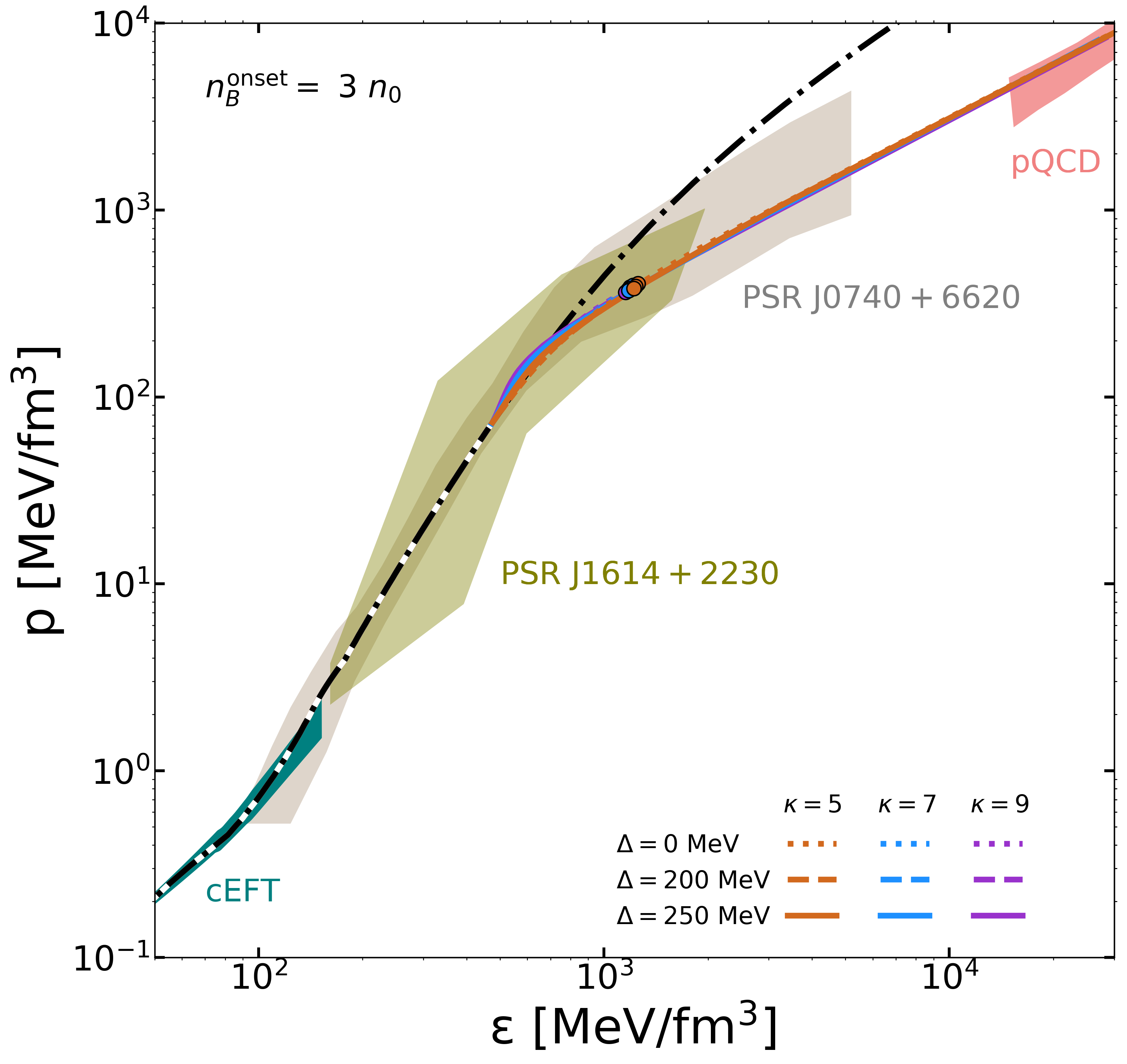}};
        \begin{scope}[x={(main2.south east)}, y={(main2.north west)}]
            \coordinate (r2SW) at (0.41,0.555);
            \coordinate (r2SE) at (0.555,0.555);
            \coordinate (r2NE) at (0.555,0.73);
            \coordinate (r2NW) at (0.41,0.73);
            \draw[gray, very thick] (r2SW) rectangle (r2NE);
        \end{scope}
    \end{tikzpicture}
\end{minipage}
\hfill
\begin{minipage}{0.48\textwidth}
    \centering
    \begin{tikzpicture}[remember picture]
        \node[anchor=south west, inner sep=0] (zoom2)
        {\includegraphics[width=\linewidth]{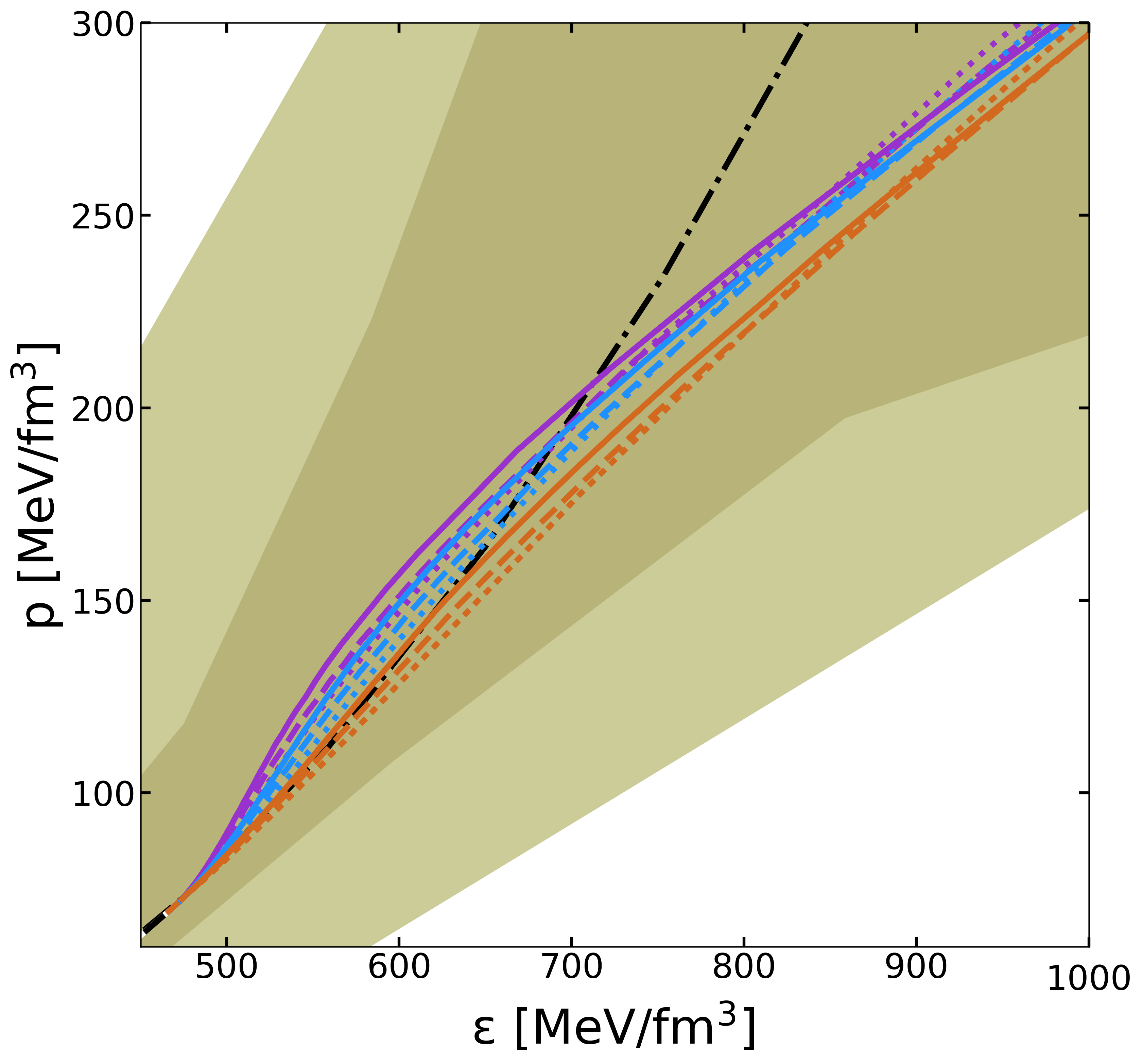}};
        \coordinate (z2NW) at ($(zoom2.north west)+(3em,-0.5em)$);
        \coordinate (z2NE) at ($(zoom2.north east)+(-1.1em,-0.5em)$);
        \coordinate (z2SW) at ($(zoom2.south west)+(3.1em,2.6em)$);
        \coordinate (z2SE) at ($(zoom2.south east)+(-1.1em,2.6em)$);
    \end{tikzpicture}
\end{minipage}

\begin{tikzpicture}[remember picture, overlay]
    \draw[gray, thick] (r1NW) -- (z1NW);
    \draw[gray, thick] (r1NE) -- (z1NE);
    \draw[gray, thick] (r1SW) -- (z1SW);
    \draw[gray, thick] (r1SE) -- (z1SE);
    \draw[gray, thick] (r2NW) -- (z2NW);
    \draw[gray, thick] (r2NE) -- (z2NE);
    \draw[gray, thick] (r2SW) -- (z2SW);
    \draw[gray, thick] (r2SE) -- (z2SE);
\end{tikzpicture}

\caption{
Pressure $p$ of electrically neutral CSQY matter at $\beta$-equilibrium as a function of the energy density $\varepsilon$ at the onset density $n_{B}^{\rm onset}=2n_0$ (upper panels) and $n_{B}^{\rm onset}=3n_0$ (lower panels) shown without zooming the region right after the onset of CSQY matter (left panels) and with that region zoomed (right panels). The calculations are performed for several values of the exponent $\kappa$ and pairing gap $\Delta$, which are indicated in the legend. The purely hadronic EoS is shown by the black dash-dotted curve in both panels. The filled circles indicate the maximum energy densities reached in the centers of the heaviest NSs modeled with the corresponding EoSs. The shaded areas represent the results of the chiral effective field theory~\cite{Hebeler:2013nza}, perturbative QCD~\cite{Kurkela:2009gj} and the constraints extracted from the binary millisecond pulsar PSR J1614-2230~\cite{Demorest:2010bx} and PSR J0740+6620~\cite{Fonseca:2021wxt}.
}
\label{fig3}

\end{figure*}

\begin{figure*}[!]
\includegraphics[width=\columnwidth]{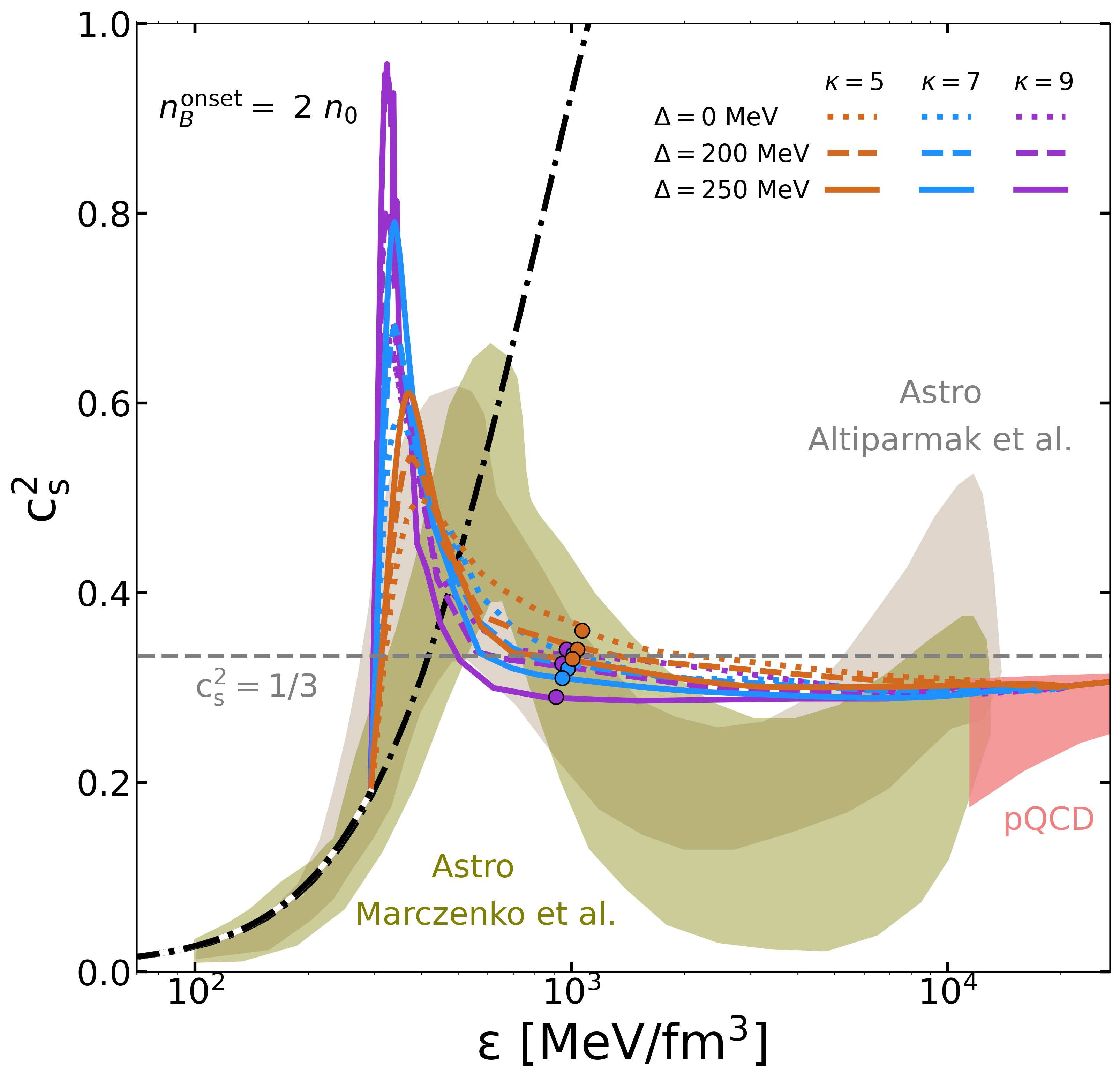}
\includegraphics[width=\columnwidth]{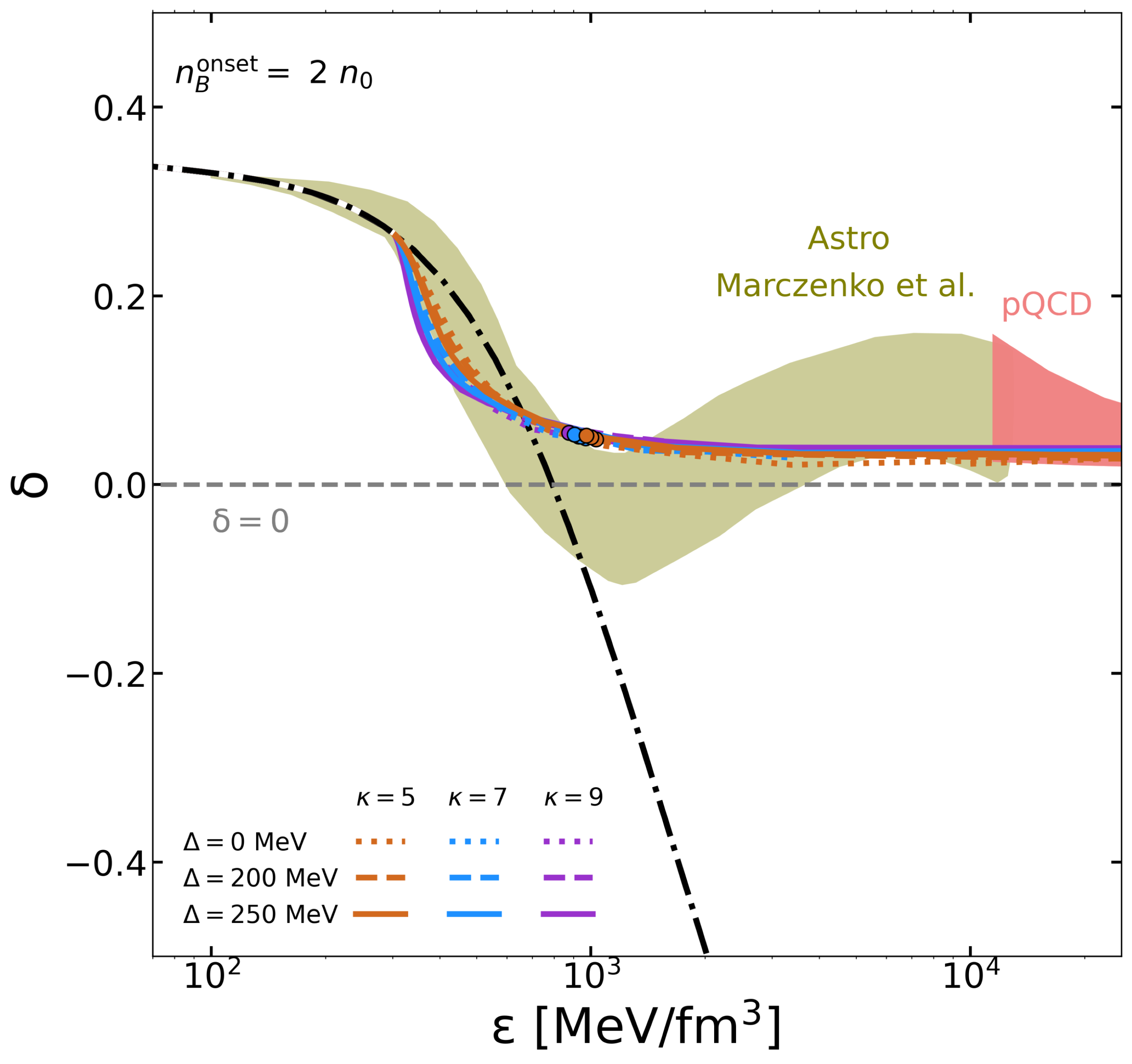}\\
\includegraphics[width=\columnwidth]{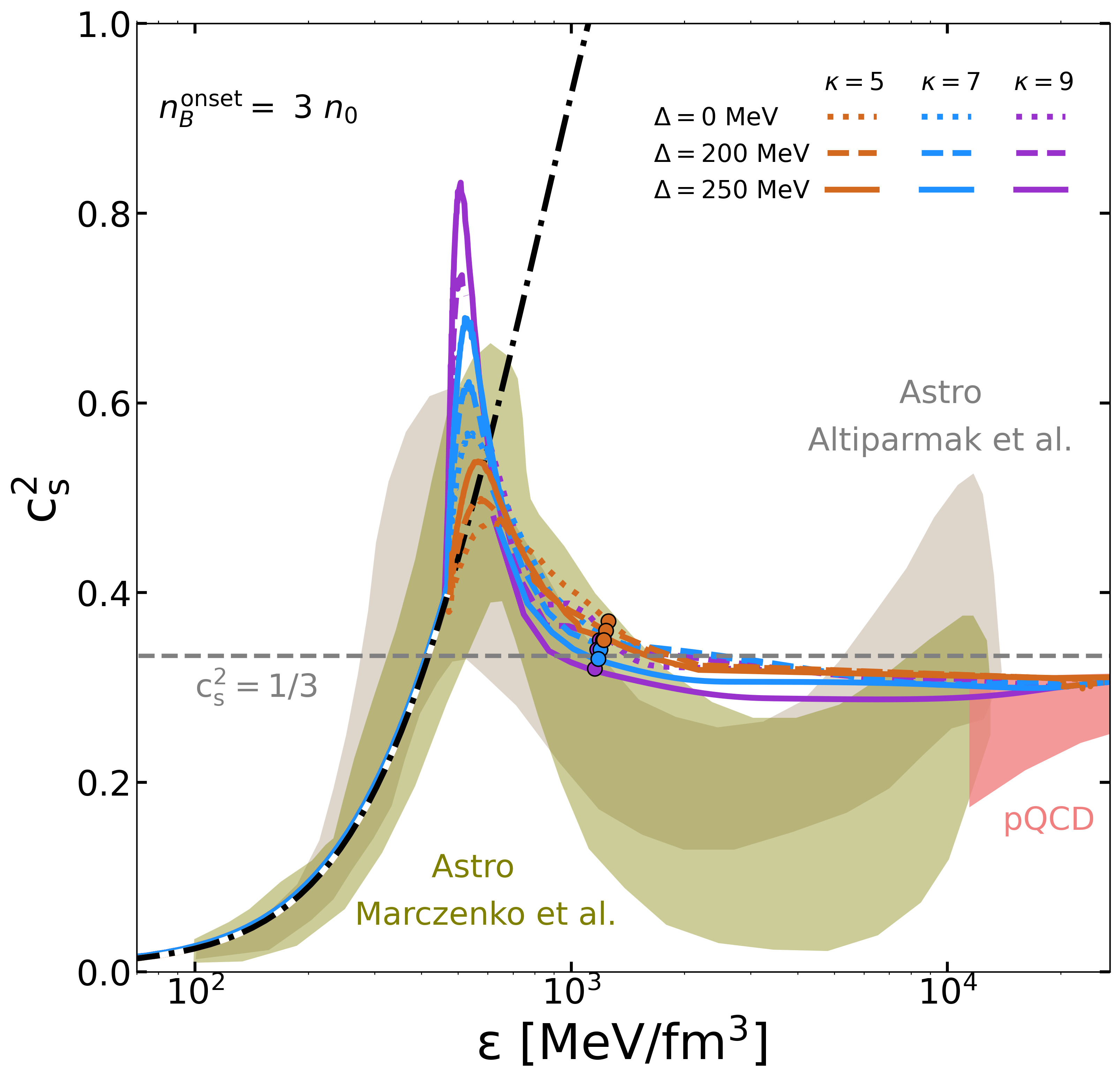}
\includegraphics[width=\columnwidth]{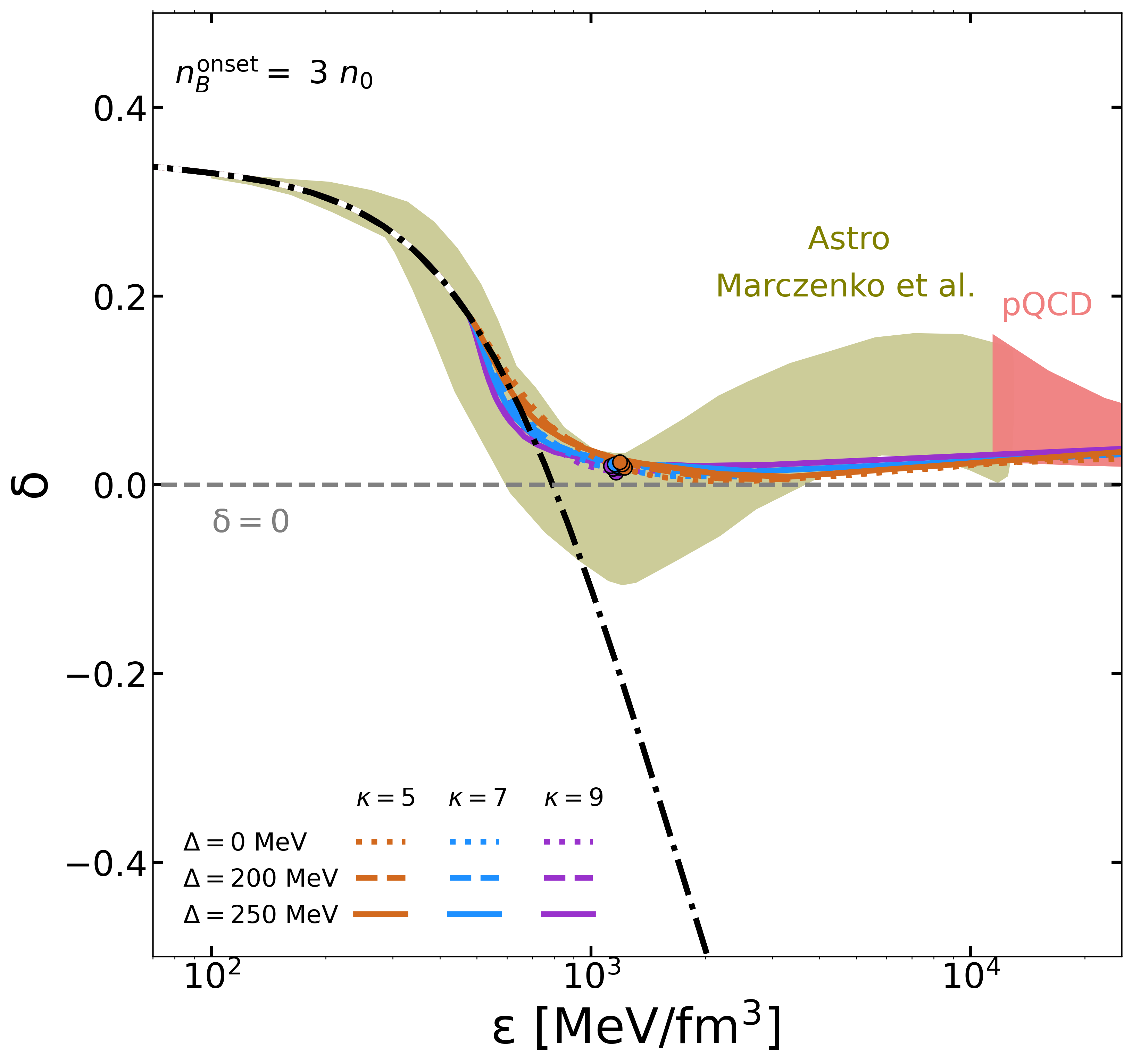}
\caption{The speed of sound $c_S^2$ (left column) and dimensionless interaction measure $\delta$ (right column) as a functions of energy density $\varepsilon$ obtained for the EoSs shown in Fig.~\ref{fig3}. The filled circles indicate the maximum energy densities reached in the centers of the heaviest NSs modeled with the corresponding EoSs. The shaded areas represent the constraints from Ref.~\cite{Altiparmak:2022bke,Marczenko:2022jhl} and perturbative QCD~\cite{Fraga:2013qra}. The idea behind the shaded regions will be explained in the text.}
\label{fig:cs_delta}
\end{figure*}
%

\subsection{Equation of state}
\label{subsec4A}

With the model parameters fixed, the EoS of CSQY matter corresponding to the conditions in the NS interior can be constructed as outlined in Sec.~\ref{sec3}. Fig.~\ref{fig3} shows it in the plane of the pressure versus energy density in comparison to the purely hadronic EoS discussed above. As is seen, the pressure of CSQY matter exceeds the hadronic one right after the onset density and gets smaller than the nuclear one at higher densities. This makes the developed model consistent not only with the low-density constraint obtained within the chiral effective field theory~\cite{Hebeler:2013nza}, but also with the high-density results of perturbative QCD~\cite{Kurkela:2009gj}. It is also worth mentioning that smaller onset densities lead to the stiffening of the EoS in the density range typical for interiors of NSs. Increasing the pairing gap $\Delta$ also leads to a higher pressure of CSQY matter, which manifests as a CS induced stiffening of its EoS. 

Increasing the exponent $\kappa$ leads to a similar effect. While the CS-induced stiffening of the EoS of CSQY matter is caused by the rearrangement of the quark momentum distribution (see Sec.~\ref{sec2} for discussion), the impact of $\kappa$ warrants a separate remark. It is clear from Eq. (\ref{VIII}) that the higher this exponent, the faster the nucleon momentum shell narrows down with an increasing density. This narrowing happens due to unfilling the low-momentum nucleon states and leads to a higher mean momentum of nucleons, which increases the kinetic energy part of the nucleon pressure and, consequently, stiffens the EoS of CSQY matter. In Refs.~\cite{McLerran:2018hbz,Zhao:2020dvu} this effect was observed due to changing $\Lambda$ at $\kappa=2$.

The conclusion regarding the CS-induced stiffening of the EoS of CSQY matter can also be inferred from Fig.~\ref{fig:cs_delta}, showing the speed of sound and dimensionless interaction measure. As is seen, increasing the pairing gap makes the peak of $c_S^2$ above the onset of CSQY matter more pronounced. Also, the consequent decrease of this quantity becomes steeper with the growth of $\Delta$. 
Understanding this effect requires explaining how the pairing gap amplitude $\Delta$ affects the condition of strong equilibrium.
The latter relates the chemical potentials of nucleons to the chemical potentials of quarks, which include the corresponding Fermi energies $\epsilon_{fc}$ having specific values required for ensuring the condition of strong equilibrium.
As is seen from the expression given after Eq. (\ref{II}), $\epsilon_{fc}$ is a growing function of the quark Fermi momentum $k_{fc}$ and a decreasing function of the pairing gap amplitude.
Thus, preserving the value of $\epsilon_{fc}$ needed to maintain the strong equilibrium requires a correlated growth of both $k_{fc}$ and $\Delta$.
Consequently, increasing $\Delta$ leads to higher $k_{fc}$.
At the same time, at high densities after the onset of CSQY matter, the momentum-dependent pairing gap $\Delta_{c{\bf k}}$ is Gaussian in momentum, which causes its suppression due to the increase of the Fermi momentum.
This suppression drives the single particle energy of quarks toward the conformal regime and, consequently, causes a faster convergence of the speed of sound to the conformal value manifested by steepening of $c_S^2$ after its peak. 
The described behavior of $c_S^2$ is reflected in the dimensionless interaction measure, which undergoes a significant decrease at the onset of CSQY matter. Remarkably, the speed of sound remains superconformal at most of the densities typical for the NS interiors, which is another manifestation of the CS induced stiffening of its EoS.

Fig.~\ref{fig:cs_delta} shows that for not too large values of $\kappa\le5$, the present model of CSQY matter agrees well with the astrophysical constraints on speed of sound and dimensionless interaction measure~\cite{Marczenko:2022jhl,Altiparmak:2022bke} up to the densities reached in the centers of the heaviest NSs. These constraints were derived from a sampling of EoSs that are consistent with perturbative QCD and chiral effective field theory, while additionally incorporating observational constraints from NSs.
Higher values of the exponent $\kappa$ are disfavored by the constraints and, in the case of a softer hadronic model, can even violate causality (see Appendix~\ref{app:A}) for details.
At the same time, at small values of $\kappa$, i.e., $\kappa<4$, the stiffening of the CSQY matter EoS right above its onset is not very significant, especially at small values of the pairing gap. This can be seen from the lower panel of Fig.~\ref{fig:cs_delta}, which indicates that at $\kappa=5$ and $\Delta=0$ the slope of the speed of sound as a function of energy density remains almost the same in the vicinity of the CSQY matter onset. This allows us to conclude that $\kappa\simeq5$ is phenomenologically most interesting for modeling NSs.

\subsection{Conformal limit}
\label{subsec4B}

As shown in Fig.~\ref{fig:cs_delta}, CSQY matter asymptotically approaches the conformal limit of QCD, characterized by $c_S^2=1/3$ and $\delta=0$~\cite{Kurkela:2009gj,Fraga:2013qra,Gorda:2018gpy,Fernandez:2021jfr}. Consistent with perturbative QCD results, these conformal values of the speed of sound and the dimensionless interaction measure are approached from below and above, respectively. 

As discussed in Ref.~\cite{Blaschke:2022egm}, nonvanishing current masses of quarks and repulsive interaction among them is the physical reason of reaching the conformal limit as described above.
The high density asymptotes of $c_S^2$ and $\delta$ found within the three-flavor nonlocal NJL model confirm this conclusion~\cite{Ivanytskyi:2024zip}.
To clarify the features of the speed of sound and the dimensionless interaction measure in the present model, we analyze its high-density asymptotic behavior.
\begin{figure*}[ht]
\centering
\setkeys{Gin}{width=0.5\linewidth}
\begin{tabularx}{\linewidth}{XX}
\includegraphics{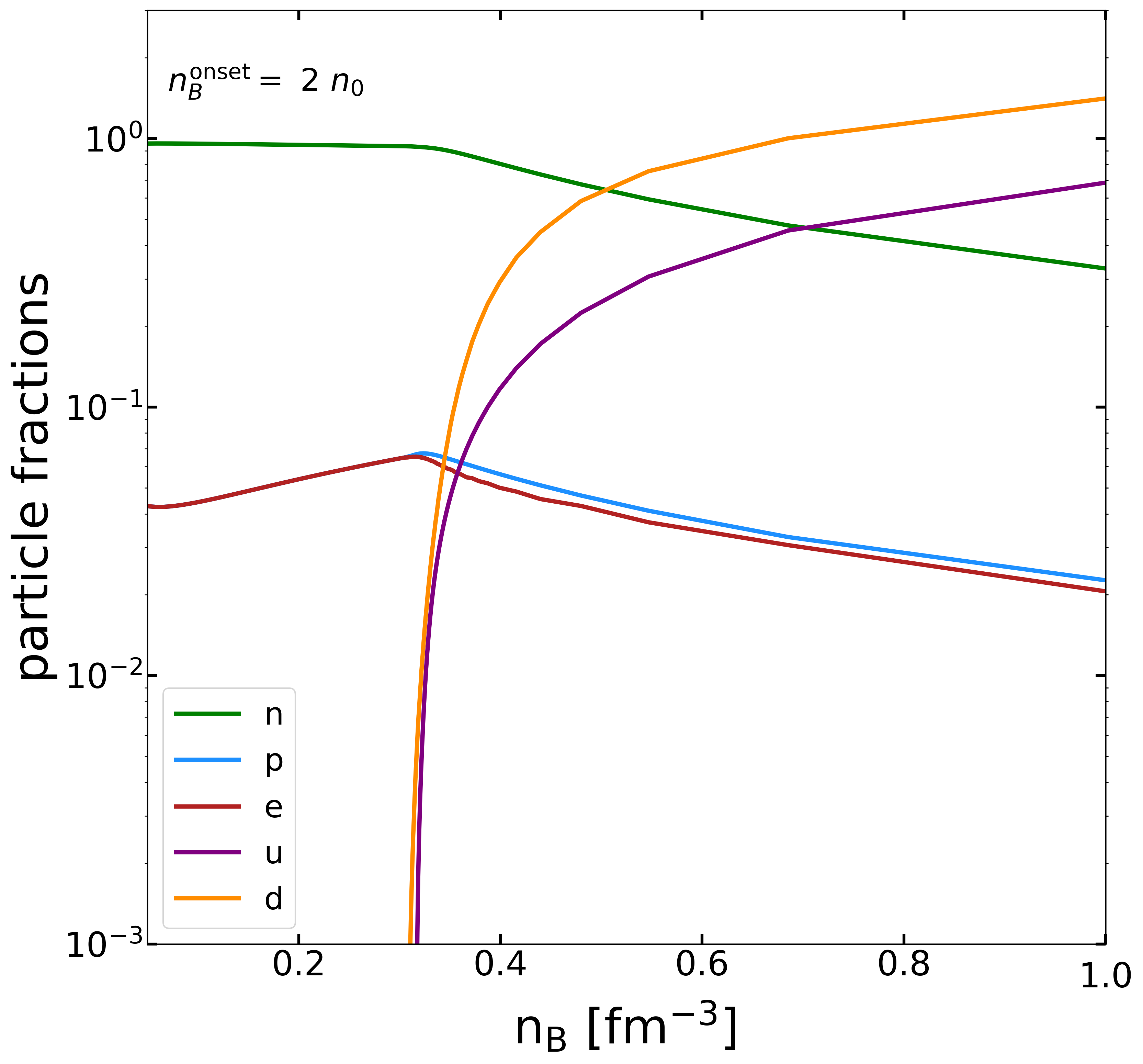}
\includegraphics{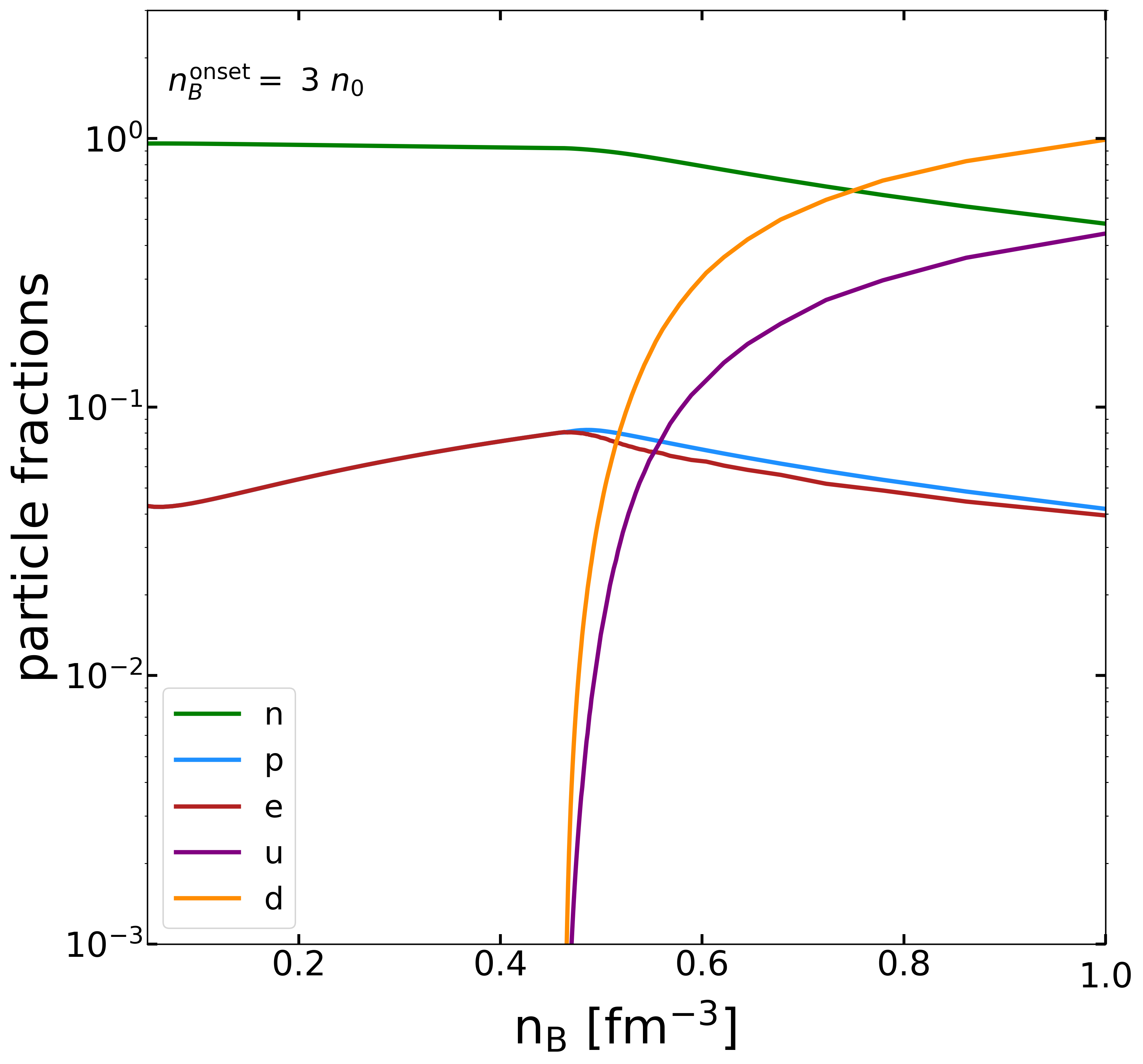}
\end{tabularx}
\caption{Particle fractions of electrically neutral CSQY matter at $\beta$-equilibrium as a function of the baryon density evaluated at the onset density $n_{B}^{\rm onset}=2n_0$, $\kappa=7$, $\Delta=200$ MeV (\textbf{left panel}) and $n_{B}^{\rm onset}=3n_0$, $\kappa=7$, $\Delta=200$ MeV (\textbf{right panel}). The particles shown are neutrons (green), protons (blue), electrons (red), u-quarks (purple), and d-quarks (orange).}
\label{fig:frac}
\end{figure*}

For the sake of simplicity, we consider the symmetric case, which preserves the qualitative conclusions of the analysis. In this regime, the baryonic chemical potential determines the thermodynamic properties of CSQY matter. A small deviation from the conformal behavior is caused by the presence of three-dimensional scales, i.e., the parameter $\Lambda_{a}$ controlling the thickness of the nucleon shell and in the symmetric case coinciding with $\Lambda$, pairing gap $\Delta$, and quark masses $m_{q}$. To account for the first of them, we notice that the vanishing of the nucleon shell thickens at high chemical potentials brings their number densities to the form $n_{a}=\Lambda^{\kappa+1} k_{a}^{2-\kappa}/\pi^{2}$, where the maximum momentum of nucleons $k_{a}$ should be replaced by its asymptote $\mu_{a}(\kappa-2)/(\kappa-3)$ found as the leading order expansion of Eq.~(\ref{XVI}). The effects of quark pairing can be perturbatively accounted for by using the corresponding correction to the quark pressure of one color-flavor state $\Delta_{fc}^2\mu_{fc}^2/4\pi^2$, where $\Delta_{fc}=(\Delta_\mu,\Delta_\mu,0)$ and $\Delta_\mu$ is the momentum-dependent pairing gap evaluated at $|{\bf k}|=\mu_B/3$ (see Section II.C of Ref.~\cite{Blaschke:2022egm} for a derivation). It is worth mentioning that this correction coincides with the pairing gap term of the well-known phenomenological parameterization of the quark matter EoS by Alford, Braby, Paris, and Reddy~\cite{Alford:2004pf} if the gap $\Delta_\mu$ is treated as a medium-independent constant. Finally, the mass correction to the quark number density can be obtained by using the leading order expansion of the number density of unpaired quarks with the quark Fermi momenta expressed through their chemical potential as $k_{fc}=\mu_{fc}-m_q^2/2\mu_{fc}$. Thus, the high-density asymptote of the quark number density becomes $n_{fc}=\mu_{fc}^3/3\pi^2-\mu_{fc}m_q^2/2\pi^2-\mu_{fc}^3\Delta_{fc}^2/\Lambda^2\pi^2$, where the quark pairing term was obtained by differentiating the corresponding pressure correction with respect to $\mu_{fc}$ and keeping the leading order contribution only. To account for the condition of color neutrality (\ref{XXVIII}), we express the quark chemical potentials as $\mu_{fc}=\mu_B/3+\mu_8\lambda_{8c}$ and notice that $\mu_8$ is small, since $\Delta_{fc}\rightarrow0$ at high densities. Consequently, linearizing Eq. (\ref{XXVIII}) in $\mu_8$, we find the later as $\mu_8=\mu_B\Delta_\mu^2/9\Lambda^2$. It is important to stress that the above decomposition of the quark chemical potential, along with the conditions of strong equilibrium, leads to $\mu_a=\mu_B$. Using the above results, the leading order expression for the baryon density of symmetric CSQY matter at high densities becomes
\begin{eqnarray}
    n_B&=&\frac{2\mu_B^3}{81\pi^2}-
    \frac{m_q^2\mu_B}{6\pi^2}-
    \frac{2\Delta_\mu^2\mu_B^3}{27\pi^2\Lambda^2}
    \nonumber\\
    \label{XXXIII}
    &+&
    \bigg{(}\frac{\kappa-3}{\kappa-2}\bigg{)}^{\kappa-2}\frac{2\Lambda_a^{\kappa+1}}{\pi^2\mu_B^{\kappa-2}}.
\end{eqnarray}
With this expression, the asymptotic behavior of the speed of sound can be found as
\begin{eqnarray}
    c_S^2&=&\frac{1}{3}\left[1-\frac{9m_q^2}{\mu_B^2}-
    \frac{4\Delta_\mu^2\mu_B^2}{9\Lambda^4}\right.
    \nonumber\\
    \label{XXXIV}
    &+&
    \left.27(\kappa+1)\bigg{(}\frac{\kappa-3}{\kappa-2}\bigg{)}^{\kappa-2}\frac{\Lambda^{\kappa+1}}{\mu_B^{\kappa+1}}\right].
    \quad
\end{eqnarray}

The asymptote of the dimensionless interaction measure $\delta=1/3-c_S^2$ is obtained using the L'H\^opital's rule~\cite{Ivanytskyi:2024zip,Ivanytskyi:2025cnn}.

It follows from Eq. (\ref{XXXIV}) that the quark mass correction in the high density asymptote of the speed of sound is dominant compared to the pairing gap and hadron shell corrections if $\kappa>1$. This expression also shows that $c_S^2$ and $\delta$ approach their conformal values in agreement with the results of the perturbative QCD, i.e., from below and above, respectively. Therefore, the disappearance of the hadron shell thickness at high densities is essential to approach the conformal limit of QCD, in agreement with perturbative calculations.

\subsection{Chemical composition}
\label{subsec4C}

Fig.~\ref{fig:frac} shows the fractions of the considered particle species as a function of baryon density. The qualitative behavior is similar to the results of Refs.~\cite{Zhao:2020dvu,Margueron:2021dtx}. Before the quark onset is reached, only protons, neutrons, and electrons are present. As the baryon density increases, the fraction of protons and electrons rises until the quark onset density is reached. At $n_B^{\rm onset}$, nucleons begin to dissolve into quarks, resulting in a substantial increase of the quark fractions. We do not discriminate between paired and unpaired quarks, only showing the species of each flavor. Clearly, the number of paired (green and red) quarks is twice the number of unpaired (blue) ones. The two panels of Fig.~\ref{fig:frac} show the particle fractions for two different values of the quark onset density. 

\begin{figure*}[th!]
\centering
\setkeys{Gin}{width=0.5\linewidth}
\begin{tabularx}{\linewidth}{XX}
\includegraphics{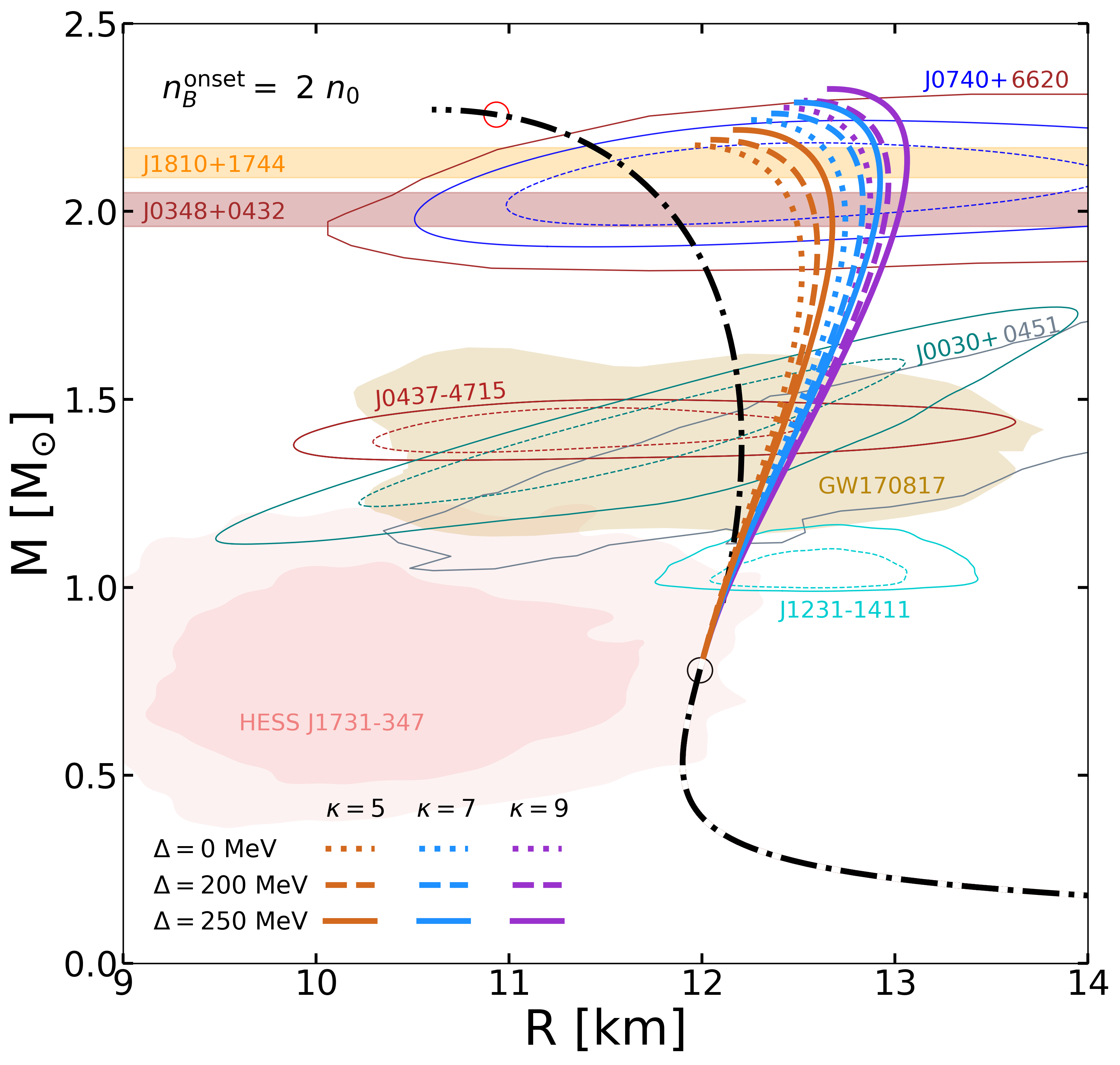}
\includegraphics[width=0.48\linewidth]{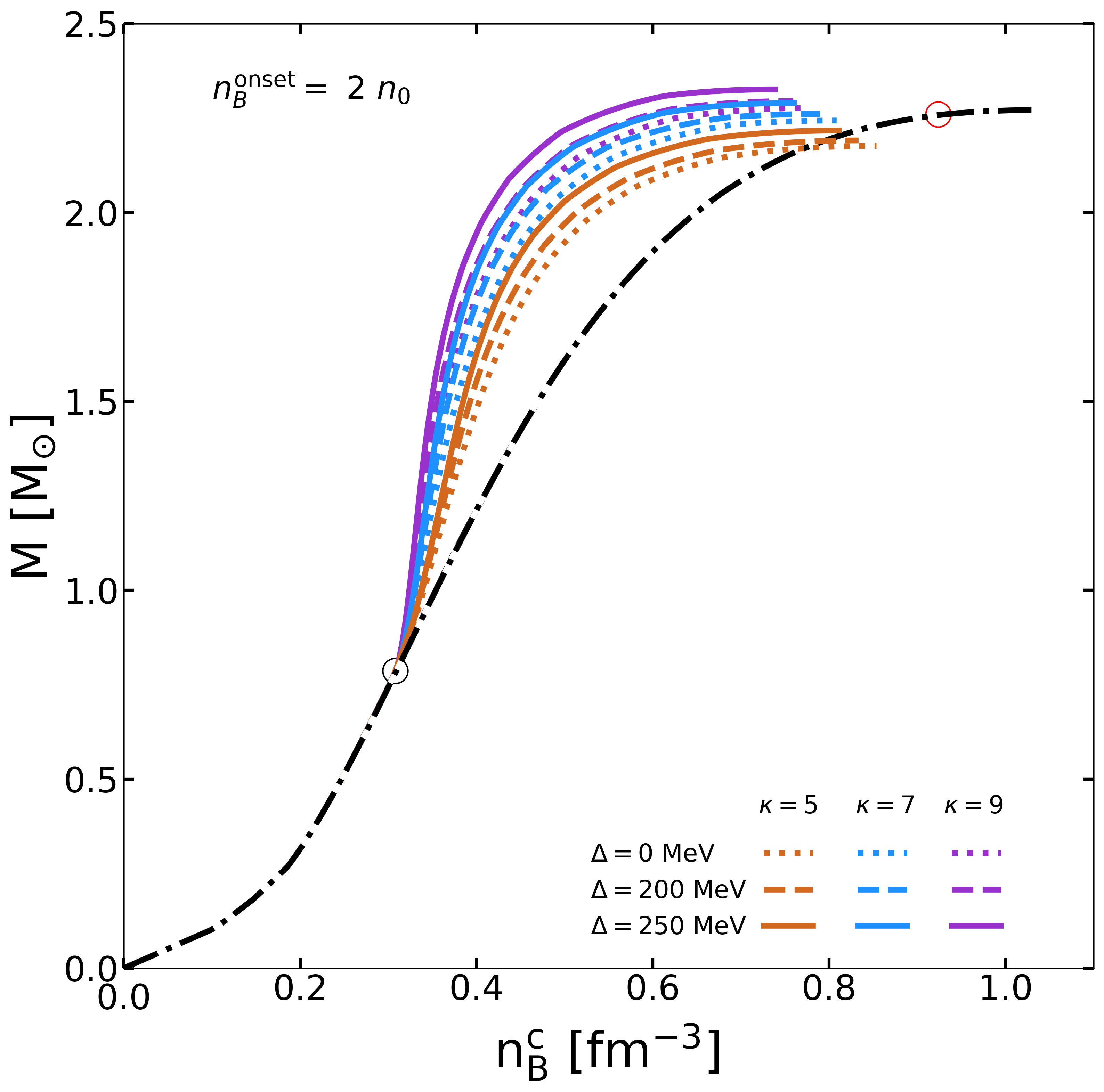}
\end{tabularx}
\begin{tabularx}{\linewidth}{XX}
\includegraphics{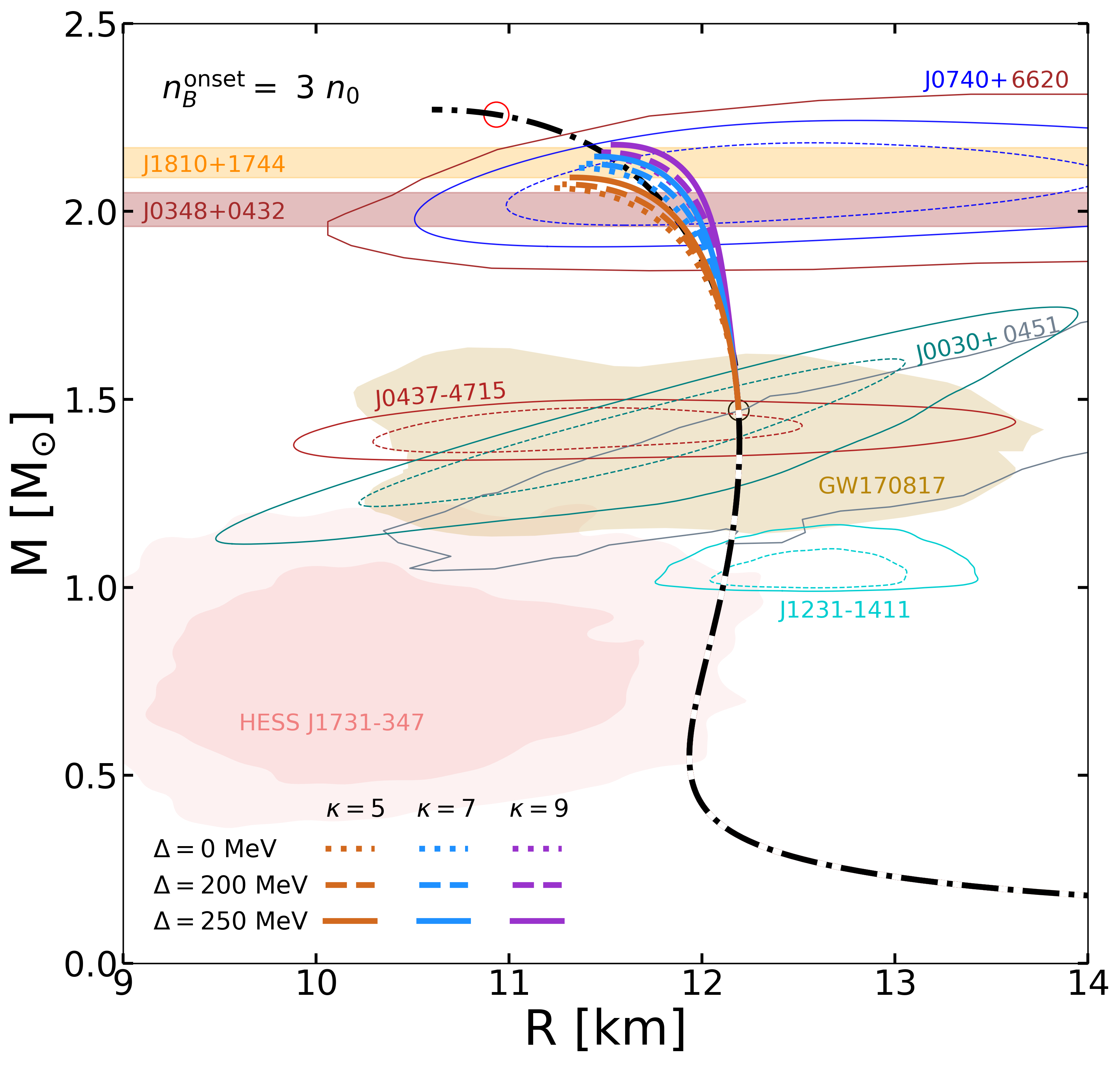}
\includegraphics[width=0.48\linewidth]{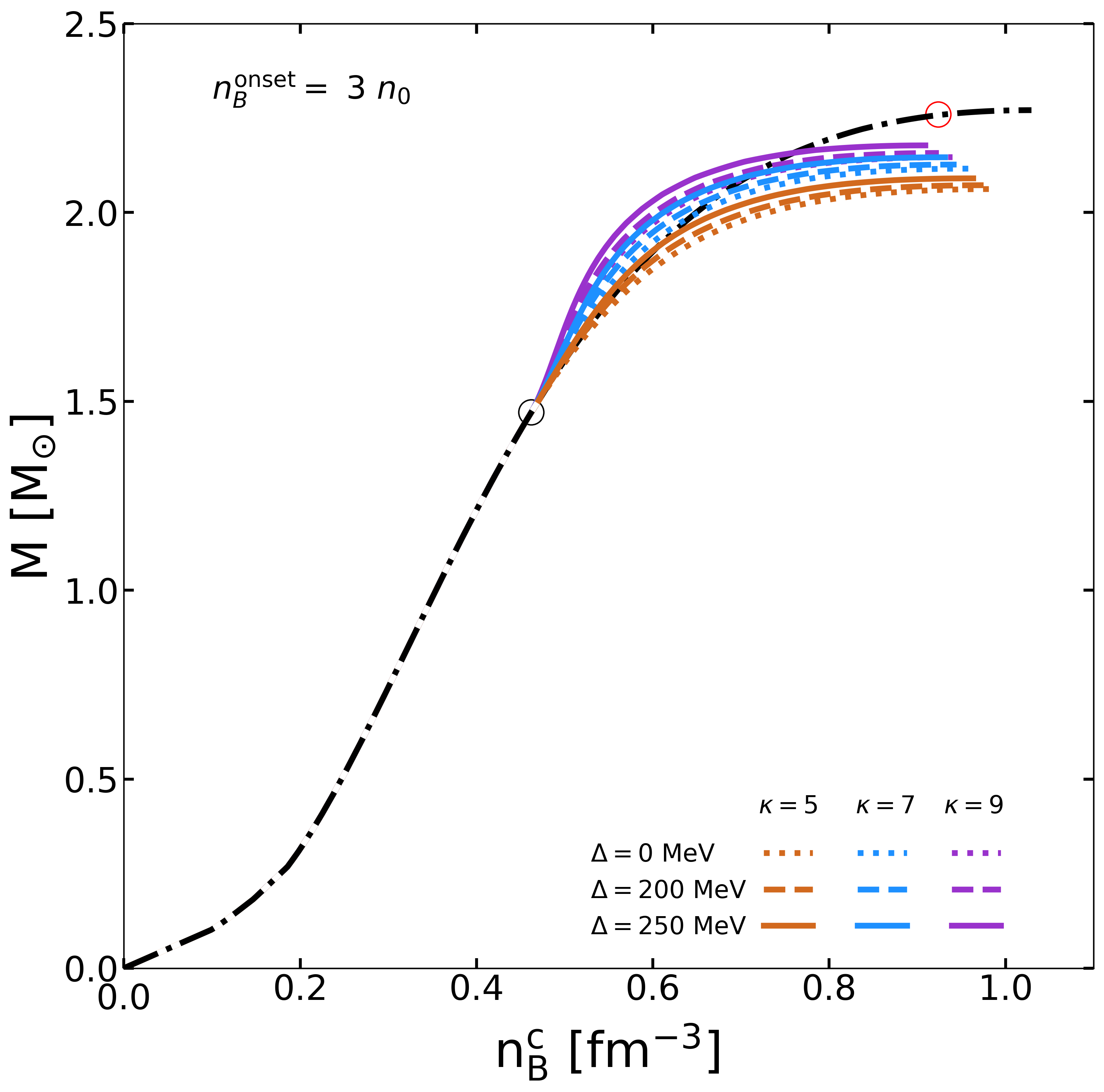}
\end{tabularx}
\caption{{\bf Left column:} $\rm M-R$ curves corresponding to EoSs shown in Fig.~\ref{fig3}. The orange and brown bands represent 1$\sigma$ constraints on the mass of PSR J1810+1744~\cite{Romani:2021xmb} and J0348+0432~\cite{Antoniadis:2013pzd}. The NICER measurements of PSR J0740+6620 are depicted with blue dotted (68 $\%$  CL) and blue solid (95 $\%$  CL)~\cite{Salmi:2024aum} as well as brown~\cite{Dittmann:2024mbo} contours. The measurements for PSR J0437-4715 are shown with firebrick dotted (68 $\%$  CL) and firebrick solid (95 $\%$  CL) contours~\cite{Choudhury:2024xbk}. The newly reported NICER data for PSR J0030+0451 are included as teal dotted (68 $\%$  CL) and teal solid (95 $\%$  CL)~\cite{Vinciguerra:2023qxq} in addition to the gray~\cite{Miller:2019cac} contour. Another recent set of data for PSR J1231-1411 is depicted with cyan dotted (68 $\%$  CL) and cyan solid (95 $\%$  CL) contours~\cite{Salmi:2024bss}. In addition, data for PSR J0614-3329 are included as olive dotted (68 $\%$  CL) and olive solid (95 $\%$  CL) contours~\cite{Mauviard:2025dmd}. LIGO-Virgo detection of GW170817 is represented in sand color~\cite{LIGOScientific:2018cki}, and the regions of 1$\sigma$ and 2$\sigma$ of HESS J1731-347 are shown in light coral~\cite {Doroshenko2022}. {\bf Right column:} Gravitational mass as a function of the central baryon density of the star corresponding to the EoSs shown in Fig.~\ref{fig3}. The hadronic M-R curves without an onset of QY matter are illustrated by the black dash-dotted curve in all panels. The red open circle indicates the causality limit for the considered hadronic EoS.}
\label{fig:M_R}
\end{figure*}
The particle abundances strongly affect a star’s thermal evolution and neutrino emission~\cite{Page:2005fq,Avila:2023rzj}. Among the various processes occurring in the NS interior, the direct Urca (DU) process, corresponding to direct and inverse $\beta$-decays of neutrons in nuclear matter, is considered as the most efficient cooling mechanism. The onset of this process is determined by the proton fraction, which, once sufficiently large, leads to enhanced neutrino emission and rapid stellar cooling. In the present model, the maximum proton fraction is determined solely by the hadronic EoS. However, as shown by~\citet{Grigorian:2004jq,Sagun:2023rzp}, quark DU processes can significantly affect the cooling behavior and highlight the role of CS in NS thermal evolution. We leave the detailed analysis of this aspect for the following study. 
\begin{figure*}[th!]
\centering
\setkeys{Gin}{width=0.5\linewidth}
\begin{tabularx}{\linewidth}{XX}
\includegraphics{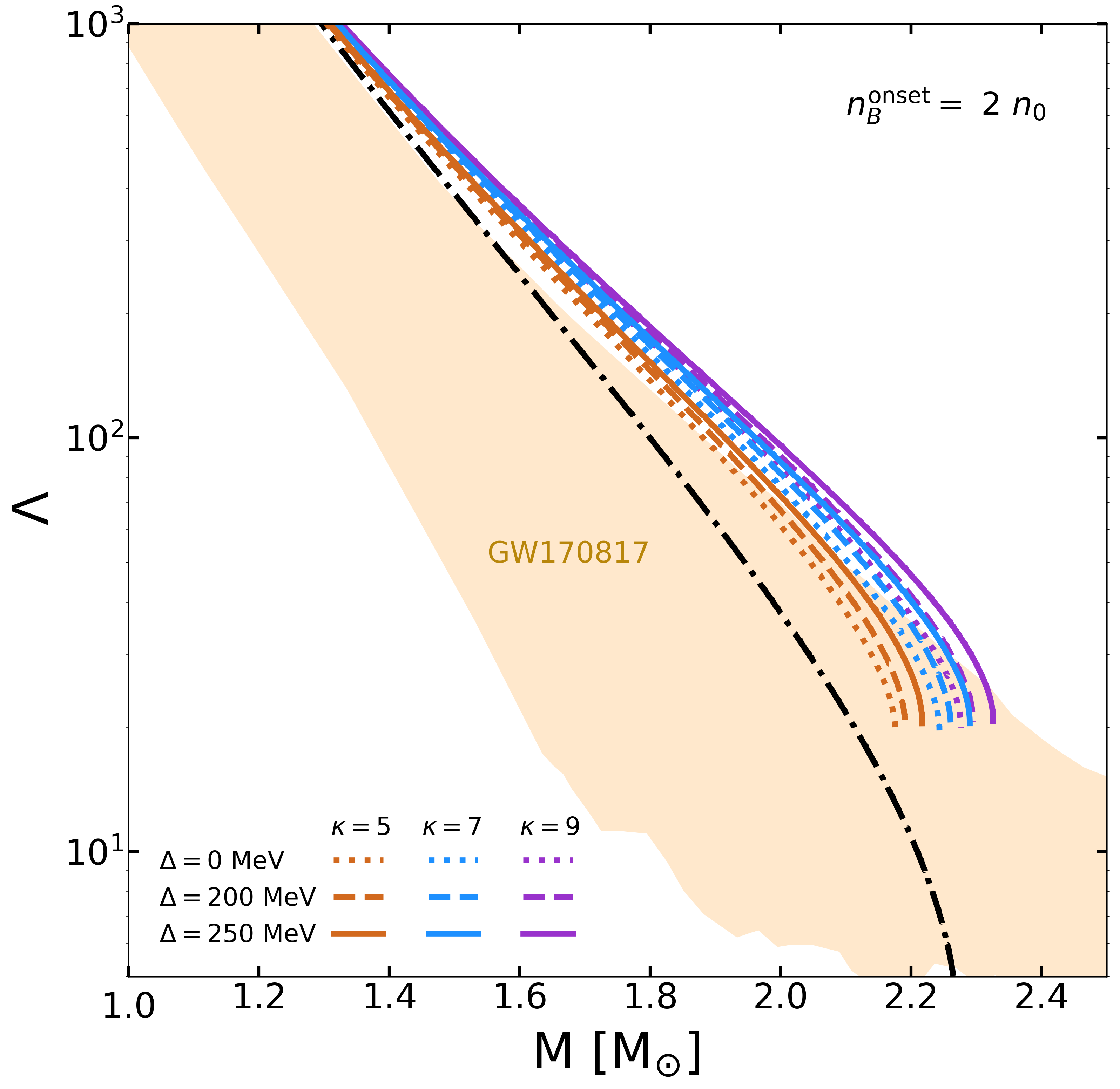}
\includegraphics{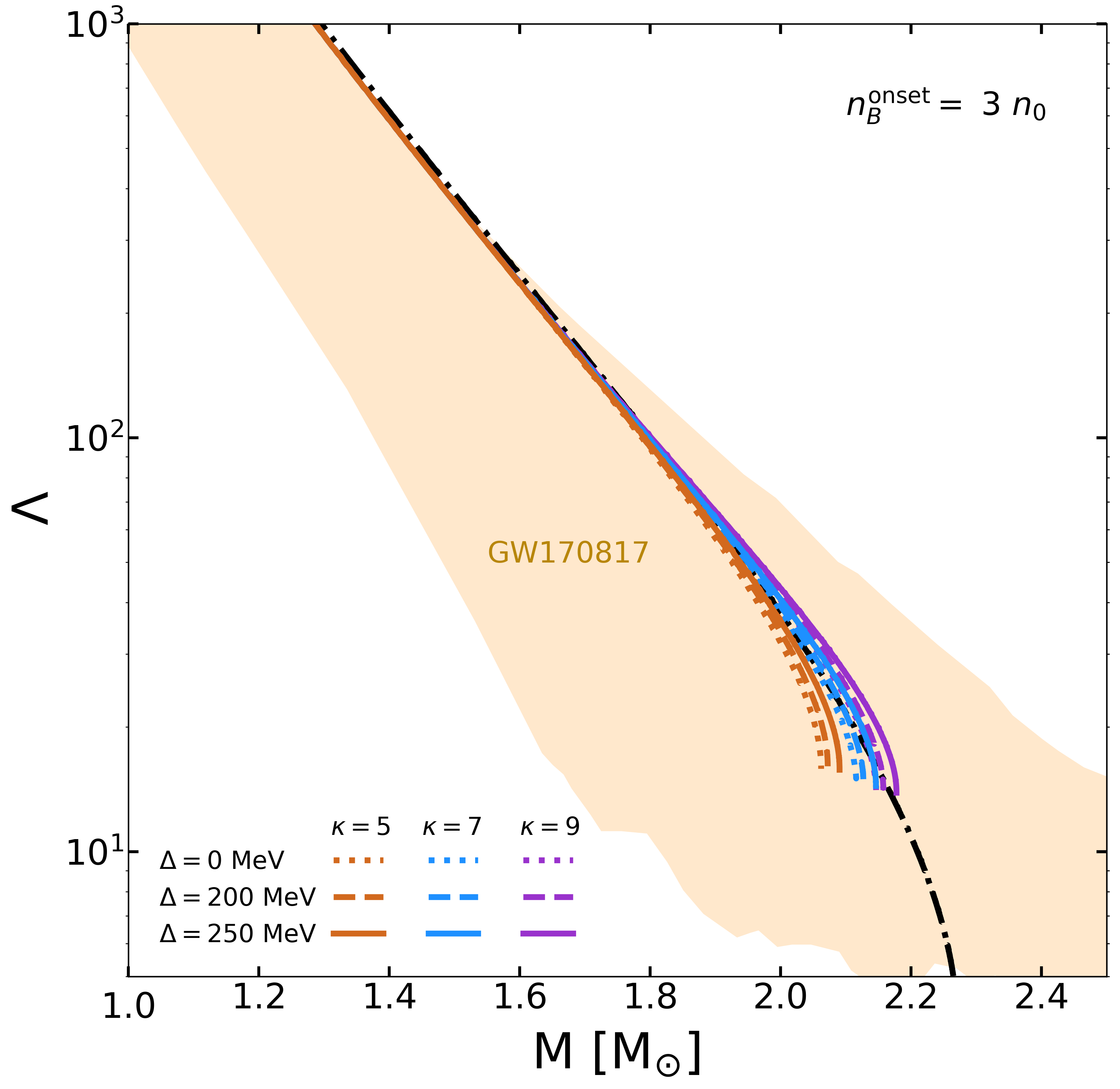}
\end{tabularx}
\caption{The tidal deformability as a function of the gravitational mass corresponding to the same EoSs as in Fig.~\ref{fig3}. The orange region depicts the constraint obtained from  GW170817~\cite{LIGOScientific:2018cki}. The purely hadronic configurations without an onset of QY matter are illustrated by the black dash-dotted curve in both panels.}
\label{fig:L_M}
\end{figure*}
%

\subsection{Color-superconducting quarkyonic stars}
\label{subsec4D}

Solving the Tolman–Oppenheimer–Volkoff equations~\cite{Tolman:1939jz,Oppenheimer:1939ne}, we obtain static, spherically symmetric equilibrium configurations of NSs with CSQY cores. Fig.~\ref{fig:M_R} presents the mass–radius relation (see the left column) and the mass–central baryon density relation (see the right column). The purely hadronic sequences are shown by the black dash-dotted curves. The onset of configurations containing CSQY cores is indicated by the open black circles, with colored curves that emerge from this point, indicating the CSQY branches. The open red circle denotes the onset of causality violation in the purely hadronic case. Each sequence is plotted up to the maximum mass configurations. It is seen that the onset of CSQY matter leads to higher radii and masses of NSs, resulting in a more pronounced back-bending of the mass-radius relations.

The impact of $\kappa$ and $\Delta$ on the stiffness of the EoS is also seen from the mass-radius curves (see the orange, blue, purple curves and the corresponding legend in Fig.~\ref{fig:M_R}). 

As discussed earlier, larger values of $\kappa$ and the effective pairing gap $\Delta$ systematically result in higher NS masses and radii. The strength of these effects depends on the onset density and on the stiffness of the hadronic EoS (see the left column of Fig.~\ref{fig:M_R}). The onset of CSQY matter at high densities diminishes the impact of quark degrees of freedom on NS properties, owing to the comparatively small size of the resulting CSQY cores. A similar effect arises from the stiffness of the hadronic EoS. As a result, for later onsets, the role of the hadronic EoS becomes increasingly dominant, allowing its properties to be more tightly constrained within the model. In general, our main results show a good agreement with the existing astrophysical and gravitational wave constraints. The transition to CSQY matter allows for more massive NSs, while also predicting significantly larger radii for high-mass stars containing a CSQY core. This is an important result, as pulsars heavier than $2.2~M_{\odot}$ are conceivable~\cite{Romani:2022jhd}.

In general, the proposed CSQY EoS provides a good description of the observational data. The hadronic curves in Fig.~\ref{fig:M_R} also agree well with the NS constraints. However, the used hadronic EoS does not include hyperons. Accounting for them would significantly soften the EoS and make it barely reach the two solar mass configurations (see e.g. Refs.~\cite{Yamamoto:2015lwa,Yamamoto:2017wre,Li:2019sxd} for discussion of this effect), which in turn would lead to a loss of consistency with the PSR J0740+6620 constraint~\cite{Salmi:2024aum}. As it was discussed in Refs.~\cite{Shahrbaf:2019vtf,Yamamoto:2021htv}, transition to stiff quark matter, which in this work is realized within the CSQY framework, resolves this tension of the purely hadronic scenario. All the depicted mass-radius curves agree with the PSR J0614-3329 data~\cite{Mauviard:2025dmd} at the 92\% confidence level.
The same quality of the data description is reached with respect to the HESS J1731-347 object~\cite{Doroshenko2022}. Further improvement of the agreement with the low mass constraints from PSR J0614-3329 and HESS J1731-347 requires a more elaborate hadronic EoS and a Bayesian scan over the parameters of the CSQY matter EoS, which is beyond the scope of the present work.

Fig.~\ref{fig:M_R} also demonstrates that once quark degrees of freedom appear, the NS mass increases significantly, while the central baryon density grows only slightly. This behavior correlates with the pronounced peak in the speed of sound at the onset of CSQY matter (see Fig.~\ref{fig:cs_delta} for details). 
This peak indicates a significant stiffening of NS matter, making it less compressible and, consequently, reducing the densities reached in the NS interiors.

Fig.~\ref{fig:L_M} shows the behavior of the tidal deformability $\Lambda$ as a function of the NS mass. In the case of the early onset of CSQY matter at $n_{B}^{\text{onset}} = 2n_{0}$ the larger NS radii discussed earlier lead to higher values of $\Lambda$ compared to the purely hadronic case (shown by the black dash-dotted curve), since $\Lambda \propto (R/M)^5$~\cite{Hinderer:2007mb,Damour:2009vw}. For the later onset at $n_{B}^{\text{onset}} = 3n_{0}$e have explored a separate, dynamical
signature of ﬁrst-order phase transitions: the resonant
tidal excitation of interfacial i -modes. The detection
of an interface mode in a coalescing neutron-star binary
would be a smoking-gun signature of a ﬁrst-order phase
transition, distinguishable in a single gravitational-wave
event. , the deviation from the purely hadronic curve is small, with only a slight decrease in $\Lambda$ at high masses. The results are compared with the 90\% confidence-level region inferred from the gravitational-wave event GW170817~\cite{LIGOScientific:2018cki}. 

\section{Conclusions}
\label{sec5}

For the first time, we propose a model of QY matter that incorporates the effects of CS in two-flavor quark matter and investigate its implications for NS phenomenology.
The most important difference compared to the earlier models of QY matter corresponds to the modification of the single particle energies and distribution functions. For paired quark states, they are postulated in the Nambu-Gorkov form and explicitly include the pairing gap. Given the exploratory character of the study, the pairing gap is not solved self-consistently but is treated as a parameter. Its momentum dependence is motivated by the running of the QCD coupling and is introduced similarly to chiral quark models with nonlocal interaction, i.e., through a momentum-dependent form factor. The used form of the form factor is chosen to provide a simultaneous onset of all color-flavor states of quarks, which is an essential element of the work. Particularly, the form factor is proportional to small momenta and vanishes according to the Gaussian scaling at high momenta.

Another modification compared to the earlier models of QY matter is related to the parameterization of the momentum shell on nucleons. The thickness of the latter is chosen in a flexible form controlled by a decreasing power-law dependence on the maximum momentum of nucleons and vanishes at asymptotically high densities. This provides vanishing of hadrons at asymptotically high densities, which reflects their dissociation to quarks driven by the asymptotic freedom of QCD. We demonstrate that a proper behavior of the nucleon shell thickness is required for reaching the conformal limit of QCD in agreement with the perturbative calculations, i.e., approaching from below for the speed of sound and from above for the dimensionless interaction measure.

To develop a model of CSQY matter applicable at arbitrary baryon densities and isospin asymmetries, we propose a modified version of the meta-model of the nuclear matter EoS, which explicitly includes the terms of kinetic energy of nucleons. Accounting for these terms is important for respecting the condition of strong equilibrium at high densities, when the interaction energy of nucleons gets small due to the vanishing of their number density, and the main contribution to the chemical potentials of nucleons is generated by their kinetic energy. Formulating these conditions in the case of CSQY is an important new element of the work. This allows us to apply the developed model of CSQY matter for constructing the EoS of QCD matter at arbitrary baryon densities and isospin asymmetries. The EoS has three free parameters, which are not excluded by the requirements of strong equilibrium, i.e., the onset density of CSQY matter, the pairing gap amplitude, and the exponential controlling vanishing of the nucleon shell.

By supplementing the developed EoS with the conditions of $\beta$-equilibrium and electric neutrality, provided by a proper amount of electrons, we obtain the EoS of NS matter and apply it to modeling these astrophysical objects. The model is also shown to be consistent with the constraints of the $\chi$EFT and perturbative QCD. We consider two values of the onset density of CSQY matter, i.e., two and three saturation densities, and several values of the pairing gap and the exponential controlling behavior of the nucleon shell. While lowering the onset density stiffens the EoS of CSQY matter, increasing each of the latter two parameters makes the characteristic peak in the behavior of its squared speed of sound more pronounced.
In other words, strengthening the effects of CS and diminishing the role of hadrons stiffen the NS EoS. This leads to higher masses, radii, and tidal deformability of NSs.

The developed model of NS matter with CSQY cores provides a good agreement with the observational data. At the same time, a more systematic analysis of its parameter space is required. The model does not account for the effects of spontaneous breaking and dynamical restoration of chiral symmetry, and the effects of vector repulsion at high densities.
Studying them along with using a self-consistently defined pairing gap deserves a separate study.

\section*{Acknowledgments}

The authors thank S. Reddy, A. Schmitt, and D. Blaschke for the fruitful discussions and valuable comments. C.G. and I.L. express their gratitude to the Funda\c c\~ao para a Ci\^encia e Tecnologia (FCT), Portugal, for providing financial support to the Center for Astrophysics and Gravitation (CENTRA/IST/ULisboa) through Grant Project No. UIDB/00099/2025. C.G. also acknowledges the Funda\c c\~ao para a Ci\^encia e Tecnologia (FCT), Portugal, through the IDPASC PT-CERN program with the No. PRT/BD/154664/2022. This work of O.I. was performed within the program Excellence Initiative--Research University of the University of Wrocław of the Ministry of Education and Science and received funding from the Polish National Science Center under Grant No. 2021/43/P/ST2/03319. V.S. gratefully acknowledges support from the UKRI-funded ``The next-generation gravitational-wave observatory network'' project (Grant No. ST/Y004248/1). This work was produced with the support of INCD and funded by FCT I.P. under Advanced Computing Project No. 2023.10526.CPCA.A2 with DOI identifier 10.54499/2023.10526.CPCA.A2.


\begin{figure*}[ht]
\centering

\begin{minipage}{0.48\textwidth}
    \centering
    \begin{tikzpicture}[remember picture]
        \node[anchor=south west, inner sep=0] (main1) 
        {\includegraphics[width=\linewidth]{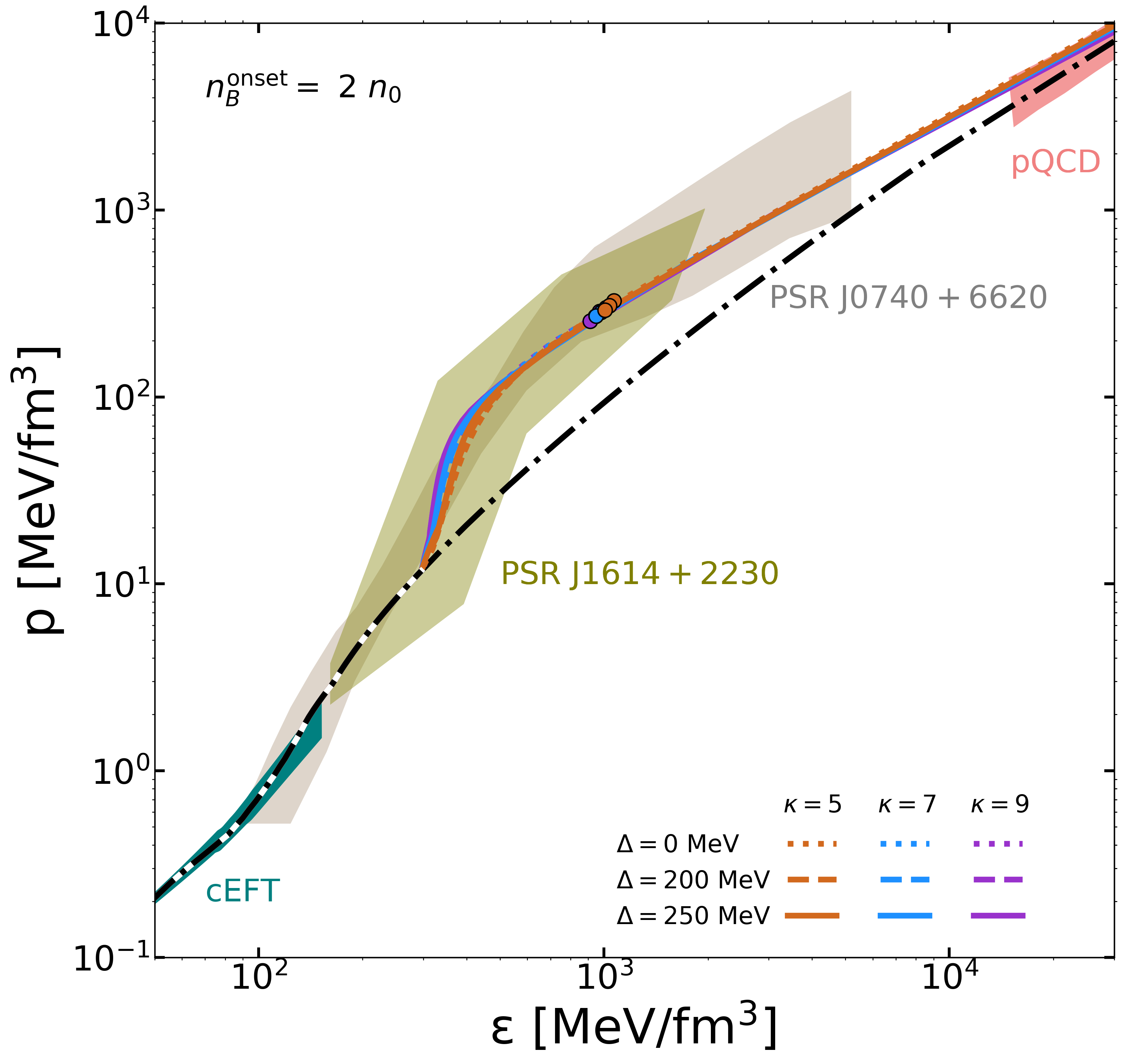}};
        \begin{scope}[x={(main1.south east)}, y={(main1.north west)}]
            \coordinate (r1SW) at (0.37,0.46);
            \coordinate (r1SE) at (0.51,0.46);
            \coordinate (r1NE) at (0.51,0.685);
            \coordinate (r1NW) at (0.37,0.685);
            \draw[gray, very thick] (r1SW) rectangle (r1NE);
        \end{scope}
    \end{tikzpicture}
\end{minipage}
\hfill
\begin{minipage}{0.48\textwidth}
    \centering
    \begin{tikzpicture}[remember picture]
        \node[anchor=south west, inner sep=0] (zoom1) 
        {\includegraphics[width=\linewidth]{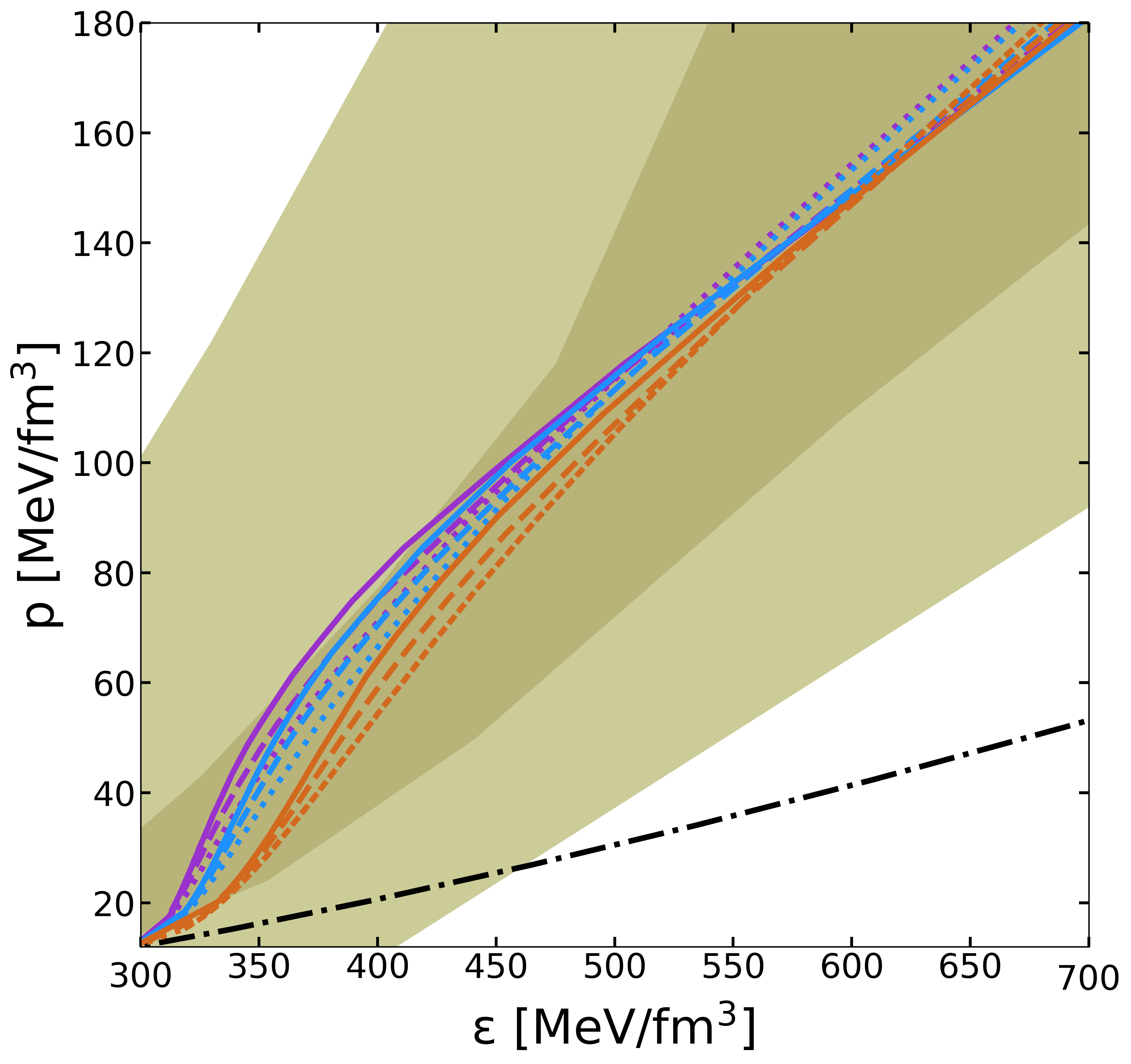}};
        \coordinate (z1NW) at ($(zoom1.north west)+(3em,-0.5em)$);
        \coordinate (z1NE) at ($(zoom1.north east)+(-1em,-0.6em)$);
        \coordinate (z1SW) at ($(zoom1.south west)+(3.2em,2.7em)$);
        \coordinate (z1SE) at ($(zoom1.south east)+(-1em,2.7em)$);
    \end{tikzpicture}
\end{minipage}

\vspace{1em}

\begin{minipage}{0.48\textwidth}
    \centering
    \begin{tikzpicture}[remember picture]
        \node[anchor=south west, inner sep=0] (main2) 
        {\includegraphics[width=\linewidth]{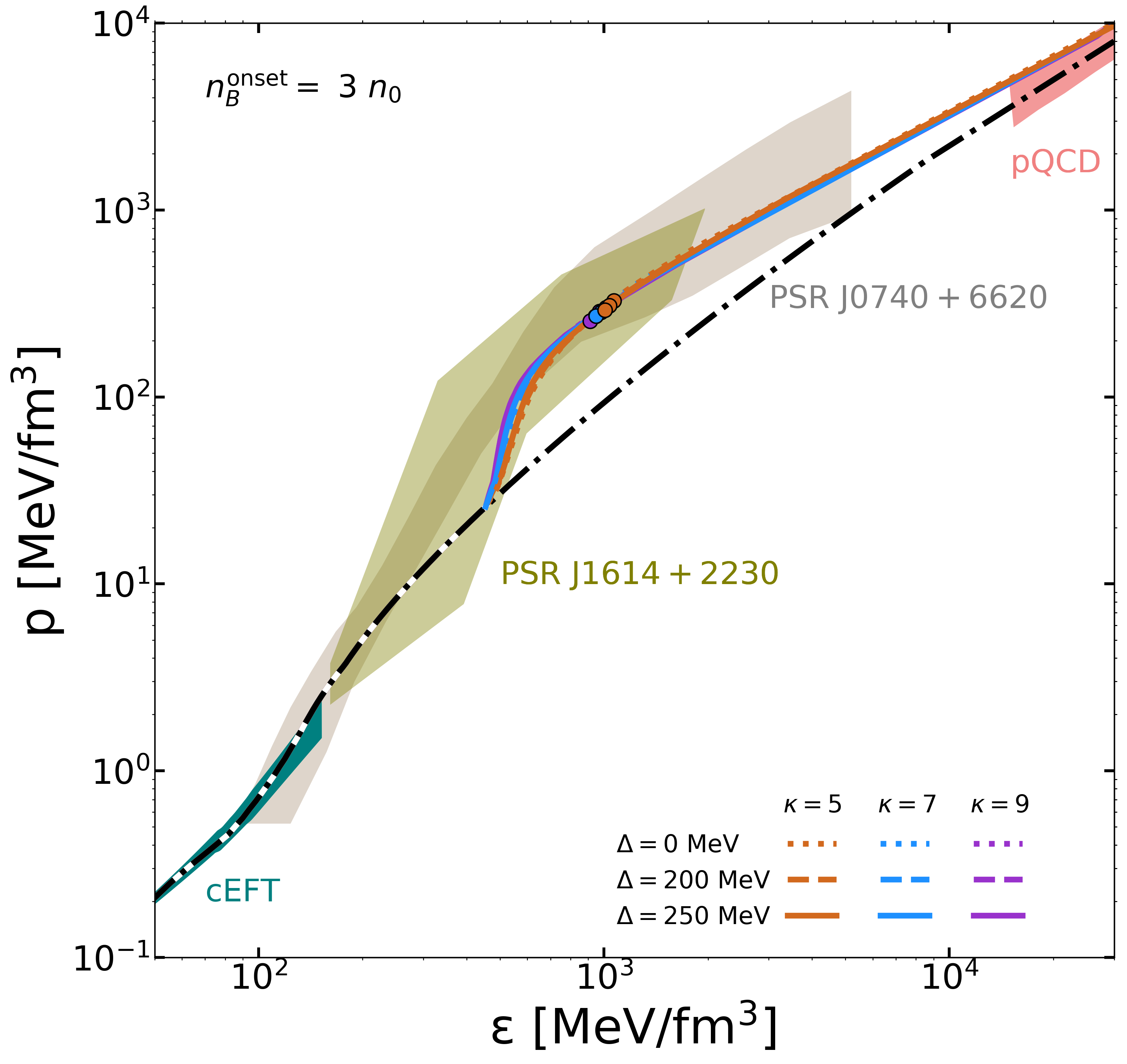}};
        \begin{scope}[x={(main2.south east)}, y={(main2.north west)}]
            \coordinate (r2SW) at (0.43,0.52);
            \coordinate (r2SE) at (0.53,0.52);
            \coordinate (r2NE) at (0.53,0.69);
            \coordinate (r2NW) at (0.43,0.69);
            \draw[gray, very thick] (r2SW) rectangle (r2NE);
        \end{scope}
    \end{tikzpicture}
\end{minipage}
\hfill
\begin{minipage}{0.48\textwidth}
    \centering
    \begin{tikzpicture}[remember picture]
        \node[anchor=south west, inner sep=0] (zoom2)
        {\includegraphics[width=\linewidth]{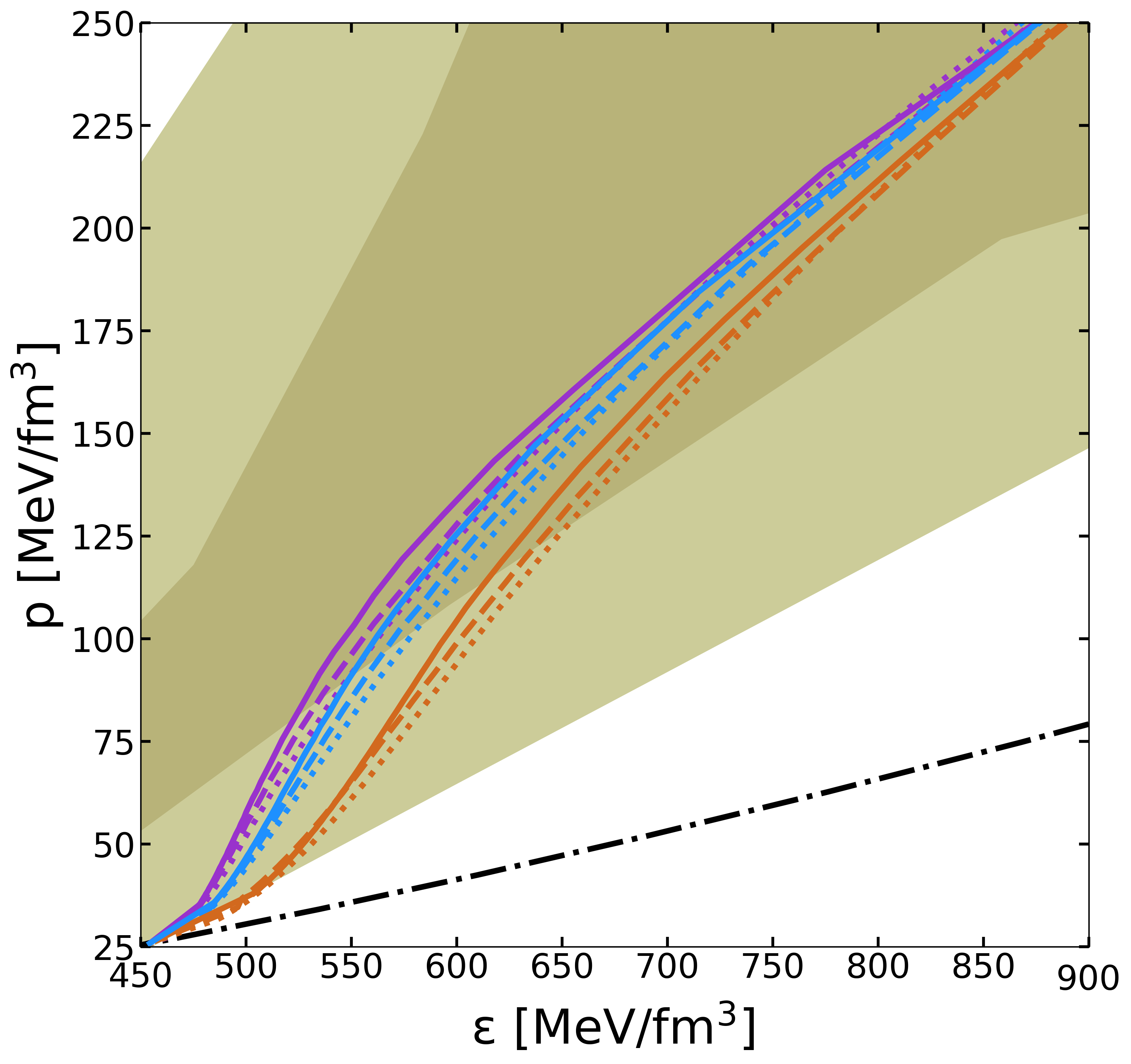}};
        \coordinate (z2NW) at ($(zoom2.north west)+(3em,-0.5em)$);
        \coordinate (z2NE) at ($(zoom2.north east)+(-1.1em,-0.5em)$);
        \coordinate (z2SW) at ($(zoom2.south west)+(3.1em,2.6em)$);
        \coordinate (z2SE) at ($(zoom2.south east)+(-1.1em,2.6em)$);
    \end{tikzpicture}
\end{minipage}

\begin{tikzpicture}[remember picture, overlay]
    \draw[gray, thick] (r1NW) -- (z1NW);
    \draw[gray, thick] (r1NE) -- (z1NE);
    \draw[gray, thick] (r1SW) -- (z1SW);
    \draw[gray, thick] (r1SE) -- (z1SE);
    \draw[gray, thick] (r2NW) -- (z2NW);
    \draw[gray, thick] (r2NE) -- (z2NE);
    \draw[gray, thick] (r2SW) -- (z2SW);
    \draw[gray, thick] (r2SE) -- (z2SE);
\end{tikzpicture}

\caption{
The same as Fig.~\ref{fig3}, but obtained with the soft nuclear EoS.}.

\label{fig:p_a}

\end{figure*}
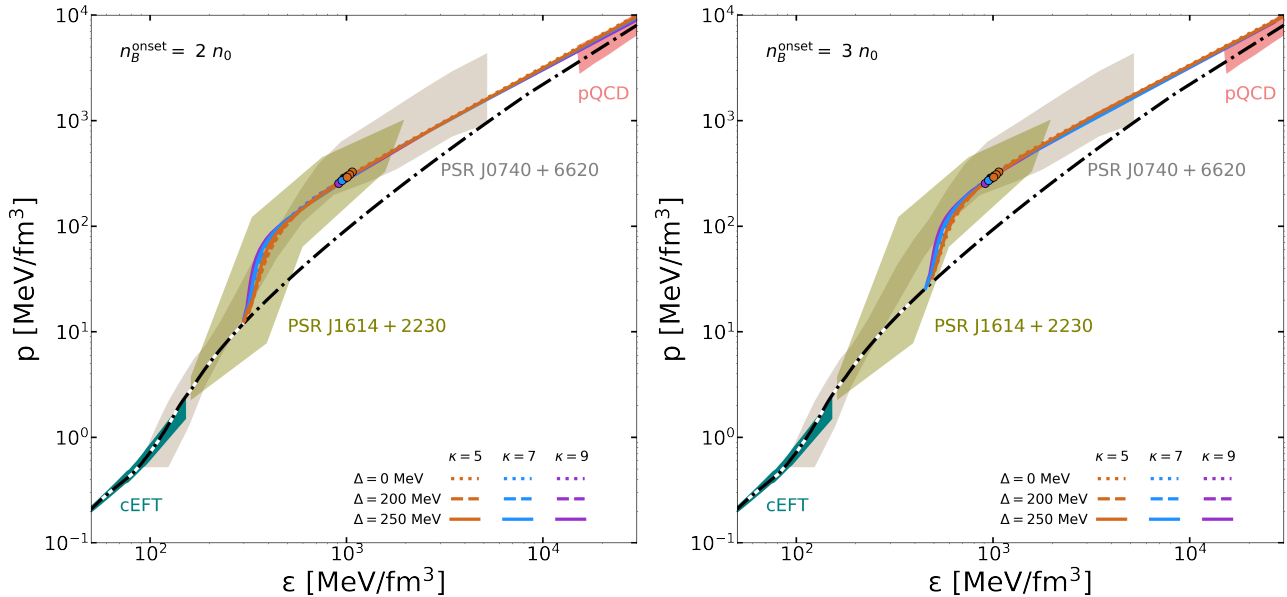

\section*{Data availability}

The data supporting the findings of this article are openly available, provided that an appropriate citation is given~\cite{gärtlein_2025_170380983}.

%
\begin{figure*}[th!]
\centering
\setkeys{Gin}{width=0.5\linewidth}
\begin{tabularx}{\linewidth}{XX}
\includegraphics{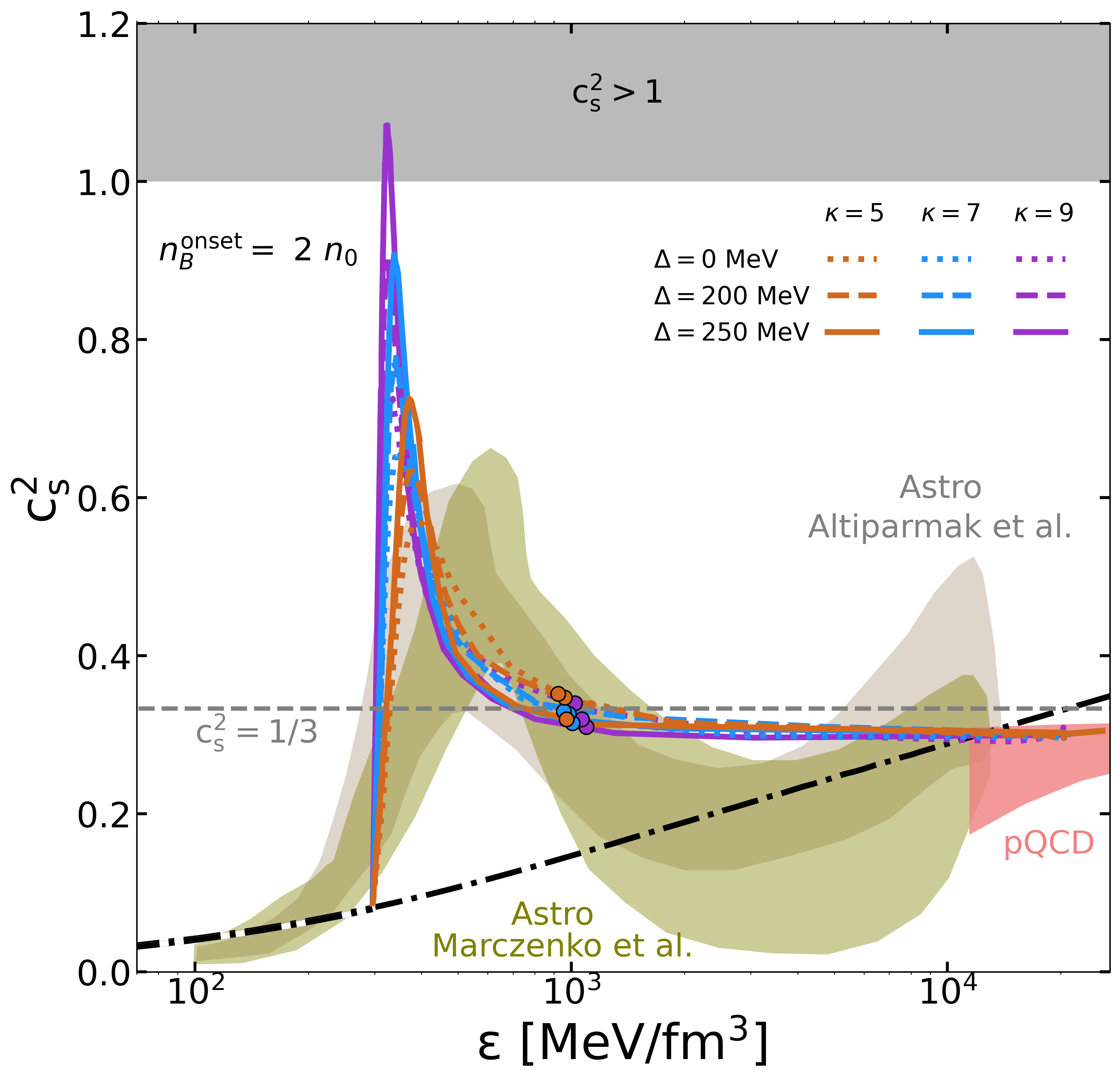}
\includegraphics{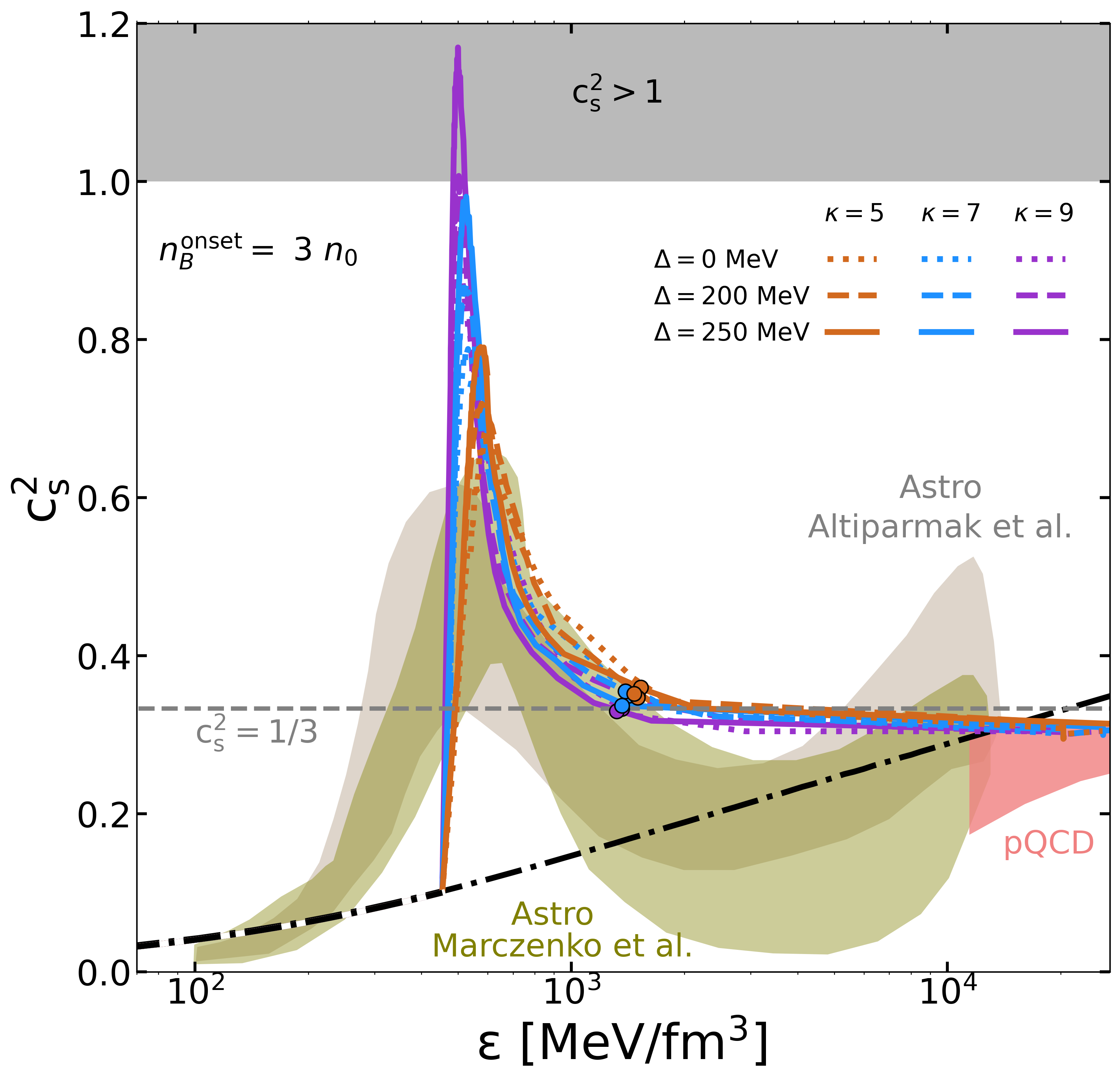}
\end{tabularx}
\caption{The same as Fig.~\ref{fig:cs_delta}, but obtained with the EoSs shown in Fig.~\ref{fig:p_a}. The gray shaded area shows the acausal region where $c_S^2>1$.}
\label{fig:c_a}
\end{figure*}
%

\begin{appendix}
    
\section{Soft hadronic EoS}
\label{app:A}

As discussed in Sec.~\ref{sec3}, we address the role of the nuclear mater EoS by considering here its softer version. The latter is obtained within the same metamodeling approach as the hadronic EoS in Sec.~\ref{sec3}, but suppressing the term corresponding to the skewness in the isoscalar interaction potential, i.e., $Q_{0}=0$ in Eq.~(\ref{XIX}), while the rest of the nuclear matter parameters remain equal to the ones in Table.~\ref{table1}. As a result, the hadronic part of the CSQY EoS significantly softens. In Fig.~\ref{fig:p_a}, the pressure as a function of energy density shows, as expected, the black dash-dotted curve to be below the heavy pulsar constraints. 

A comparison of the pressure vs. energy density figures obtained for the CSQY matter for the stiff (see Fig.~\ref{fig:cs_delta}) and soft (see Fig.~\ref{fig:p_a}) hadronic EoSs reveals a good agreement with the astrophysical data. While the soft EoS does not reproduce the heavy pulsars, the onset of quarks and the consequent stiffening of matter at higher densities provide a good agreement with the data. 

The speed of sound squared depicted in Fig.~\ref{fig:c_a} shows a pronounced peak after the onset of quarks. Interestingly, the maximum values of $c_s^2$ are increased compared to the case of a stiff hadronic EoS. Thus, the peak at the onset violates causality for $\kappa=9$ and $\Delta \neq 0$, with values of $c_s^2>1$. Therefore, the combinations of $\kappa=9$ and  $\Delta > 0$ in the case of the soft hadronic EoS are excluded from the realistic parameters of the 
CSQY EoSs. Conformality is still reached from below in all cases, as seen in Fig.~\ref{fig:c_a}. In comparison to Fig.~\ref{fig:cs_delta}, consistency with the illustrated constraints~\cite{Altiparmak:2022bke,Marczenko:2022jhl} is provided for the early onset case. 

\begin{figure*}[th!]
\centering
\setkeys{Gin}{width=0.5\linewidth}
\begin{tabularx}{\linewidth}{XX}
\includegraphics{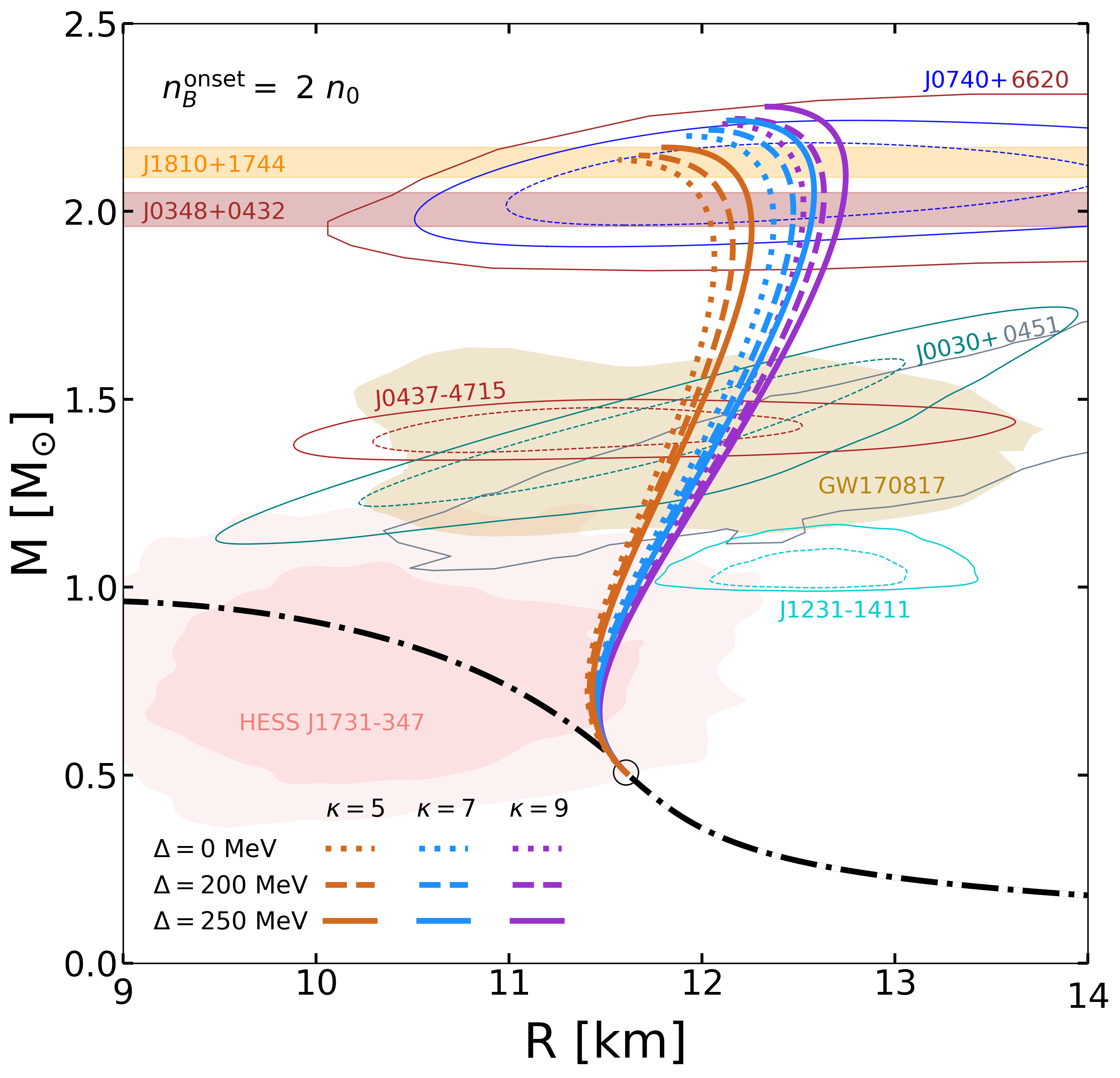}
\includegraphics{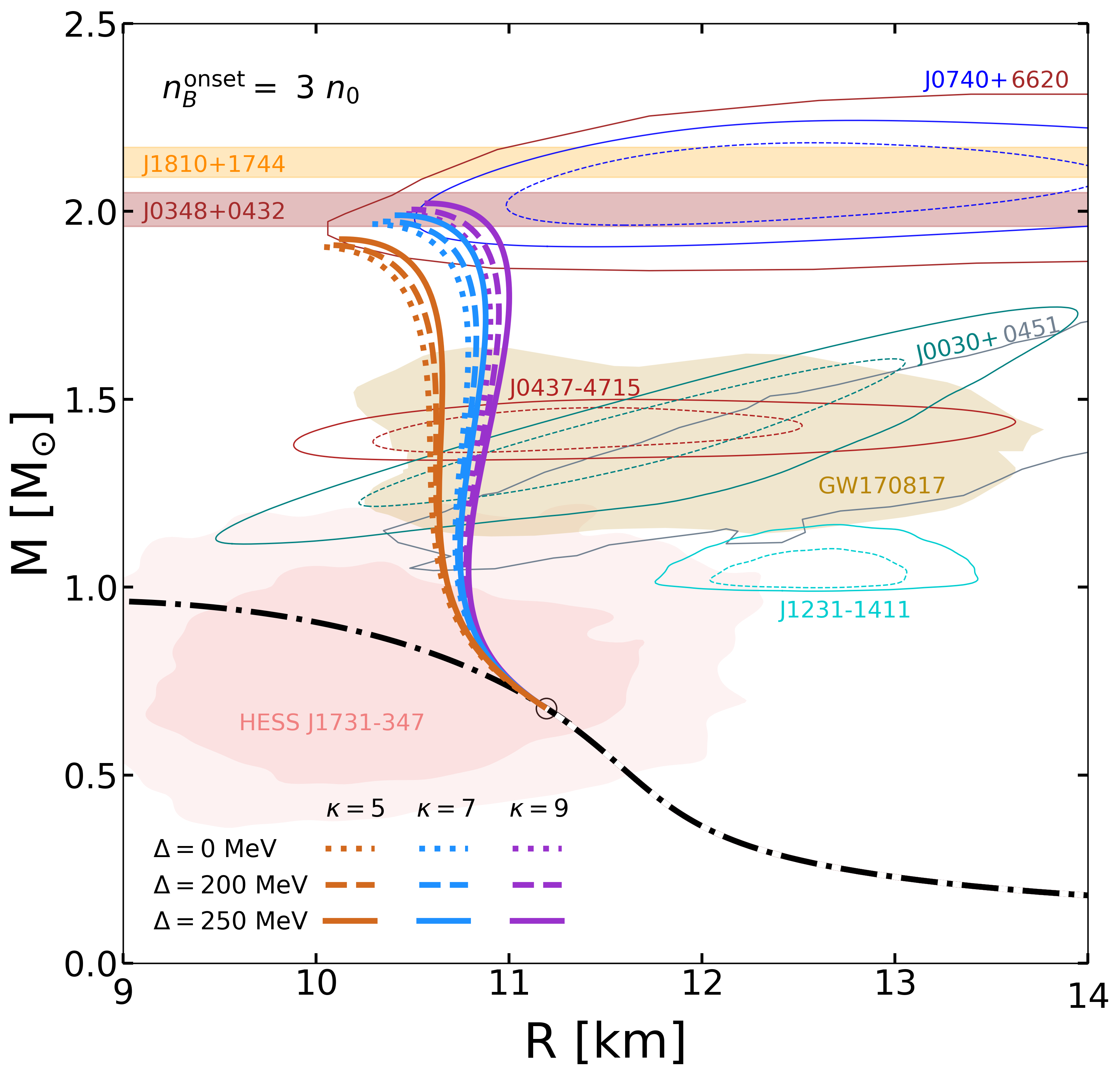}
\end{tabularx}
\caption{The same as on the left column of Fig.~\ref{fig:M_R} but obtained with the EoSs shown in Fig.~\ref{fig:p_a}.}
\label{fig:M_a}
\end{figure*}

In Fig.~\ref{fig:M_a}, the M-R curves based on the softer EoS are shown. The results are qualitatively close to the stiffer case with an early onset. The exponent $\kappa$ impacts the stiffness of the EoS and leads to more massive NSs and higher radii. In the present case, purely hadronic configurations do not reach $1\rm M_{\odot}$ and are excluded due to astrophysical constraints. In the scenario of an early onset of CSQY matter, resulting M-R curves provide a good agreement with observational data while in the case of a moderate onset, the stiffness gained from quark matter is not sufficient to satisfy the $2\rm M_{\odot}$ constraint, even though the speed of sound surpasses the speed of light at densities realized inside the NS. 

In summary, the properties of CSQY matter are strongly dependent on the underlying hadronic EoS. A late onset of CSQY matter inside the NS results in only small deviations from purely hadronic configurations. In contrast, an early onset of quark matter within the CSQY framework allows for the existence of very massive NSs, with their characteristics depending sensitively on the parameters $\kappa$ and $\Delta$.

\end{appendix}
\bibliography{references}

\end{document}